\documentclass[prl,english,superscriptaddress,preprintnumbers,reprint,footinbib,amsmath,amssymb,prl]{revtex4-2}
\usepackage{graphicx}
\usepackage[version=3]{mhchem}
\usepackage{graphicx}
\usepackage{dcolumn}
\usepackage{bm}
\usepackage{hyperref}
\hypersetup{
    colorlinks=true,
    linkcolor=blue,
    filecolor=blue,
    urlcolor=blue,
   citecolor=blue,
}
\usepackage{enumitem}
\usepackage{siunitx}
\usepackage{inputenc}
\usepackage{braket}

\begin{document}

\title{Spin-polarized Majorana zero-modes in double zigzag honeycomb nanoribbons}

\author{R. C. Bento Ribeiro}
\affiliation{Instituto de F\'{i}sica, Universidade Federal Fluminense, Av. Litor\^anea s/N, CEP: 24210-340, Niter\'oi, RJ, Brazil}
\affiliation{Centro Brasileiro de Pesquisas Físicas, Rua Dr. Xavier Sigaud, 150, Urca 22290-180, Rio de Janeiro, RJ, Brazil}
\author{J. H. Correa}
\affiliation{Universidad Tecnol\'ogica del Per\'{u},  Nathalio S\'anchez, 125, Lima, Per\'{u}}
\affiliation{Instituto de Física, Universidade de Brasília, Brasília-DF 70919-970, Brazil} 

\author{L. S. Ricco}
\affiliation{Science Institute, University of Iceland, Dunhagi-3, IS-107,Reykjavik, Iceland}


\author{A. C. Seridonio}
\affiliation{S\~ao Paulo State University (Unesp), School of Engineering, Department of Physics and Chemistry, 15385-000, Ilha Solteira-SP, Brazil}
\affiliation{S\~ao Paulo State University (Unesp), IGCE, Department of Physics, 13506-970, Rio Claro-SP, Brazil}

\author{M. S. Figueira}
\email[corresponding author: ]{figueira7255@gmail.com}
\affiliation{Instituto de F\'{i}sica, Universidade Federal Fluminense, Av. Litor\^anea s/N, CEP: 24210-340, Niter\'oi, RJ, Brazil}

\date{\today}

\begin{abstract}


We study the emergence of Majorana zero modes (MZMs) at the ends of a finite double zigzag honeycomb nanoribbon (zHNR). We show that a double zHNR geometry can host spin-polarized MZMs at its ends. We considered a minimal model composed by first nearest neighbor hopping, Rashba spin-orbit coupling (RSOC), \textit{p}-wave superconducting pairing, and an applied external magnetic field (EMF).  The energy spectrum regions with either spin up or down MZMs belong to distinct topological phase transitions characterized by their corresponding winding numbers and can be accessed by tunning the chemical potential of the nanoribbons. Hybrid systems constituted by zHNRs deposited on conventional \textit{s}-wave superconductors are potential candidates for experimentally realizing the proposal. The spin's discrimination of MZMs suggests a possible route for performing topological-conventional qubit operations using Majorana spintronics.
\end{abstract}


\maketitle


\textit{Introduction---.} 
Starting from the seminal work of Read and Green \cite{Read20} on two-dimensional \textit{p}-wave superconductors, Kitaev proposed a simplified one-dimensional (1D) toy model \cite{Kitaev2001}. In this model, unpaired MZMs appear at opposite ends of a \textit{p}-wave superconducting tight-binding chain. Remarkably, it took less than a decade~\cite{Oreg10} to understand that it could experimentally realize Kitaev's original proposal. Some setups~\cite{Lutchin10, Mourik12, Gul2018} employed hybrid devices composed of a 1D semiconductor nanowire with strong RSOC, in contact with a conventional \textit{s}-wave superconductor and under an EMF longitudinal to the nanowire. Topological protected MZMs emerge at the nanowire ends~\cite{Aguado17} when the nanowire chemical potential lies on the bulk \textit{p}-wave superconducting induced-gap.

Another kind of setup came up after the development of the epitaxially grown hybrid semiconductor-superconductor systems in which two or three facets of the hexagonal InAs nanowire core were covered by Al \cite{Krogstrup2015}. This setup is a hybrid platform that employs a naturally occurring quantum dot at the end of the nanowire as a spectrometer \cite{Clarke2017, Schuray2017, Prada2018, Ricco2018} to measure the nonlocality degree and the spin canting angles of the nonlocal MZMs \cite{Deng2016, Deng2018}. Another hybrid platform is a chain of ferromagnetic atoms aligned over a conventional \textit{s}-wave superconductor with strong RSOC ~\cite{Perge14, JeonScience2017}. In this scenario, the essential ingredients to generate MZMs at the ends of the chain are the ferromagnetic interaction between atoms that composes the chain and the RSOC induced on the chain by the superconducting substrate. We can find a helpful review of the experimental state-of-the-art on the subject in Refs. ~\cite{Flensberg2021,NatureReviewBerthold2021}.

Although the manifestation of MZMs in 1D hybrid semiconductor-superconductor nanowires has been widely explored over the last decade~\cite{Aguado17, ZhangNatCommun2019, prada2020andreev}, the emergence of topological Majorana excitations in alternative 2D honeycomb lattice platforms have only received marginal attention. Between them, we are interested in zHNRs build from Xenes graphene-like family~\cite{Aidi20},~\cite{Dutreix2014, Ma2017}; where X represents single elements from group III to group VI of the periodic table. Probably, silicene (X=Si) is the most promising candidate of this family for obtaining a zHNR geometry for hosting MZMs~\cite{Lalmi2010, Aufray2010}. Its energy spectra \cite{Liu2011} can be spin-polarized by applying an external electric field perpendicular to the zHNR sheet plane~\cite{Ezawa2012, Drummond2012, Jiao2020}, giving rise to an effective extrinsic RSOC that breaks its mirror symmetry [see Note 2 of Supplemental Material (SM)]. Silicene also presents an excellent potential to produces half-metallic transport and pure spin-current \cite{Tsai2013, Latge2015, Jiang_2019}.

Despite the spinless nature of Kitaev's work, some proposals have been exploring the spin properties of MZMs in different contexts. Jeon et al.~\cite{JeonScience2017}, employed a spin-polarized STM to distinguishes between topological MZMs and other trivial in-gap states in chains of Fe atoms deposited on top of superconducting Pb. Spin-polarization of MZMs was also accounted to investigate the Kondo effect in a quantum dot coupled to a metallic contact and a pair of MZMs~\cite{Vernek2020}, and to study the transport properties of a finite-length Majorana nanowire placed between a dot and a metallic lead~\cite{Deng2018, Schuray2020}. Moreover, the Majorana spin polarization was also employed as an alternative way of performing quantum computing operations~ \cite{LeijnsePRL2011, Flensberg2012}, allowing the transference of spin qubits between quantum dots and also realizing nontrivial two-qubit gates. However, none of these works propose a way to discriminate the spin degree of freedom of the MZMs.

\begin{figure}[t]
	\centerline{\includegraphics[width=3.0in,keepaspectratio]{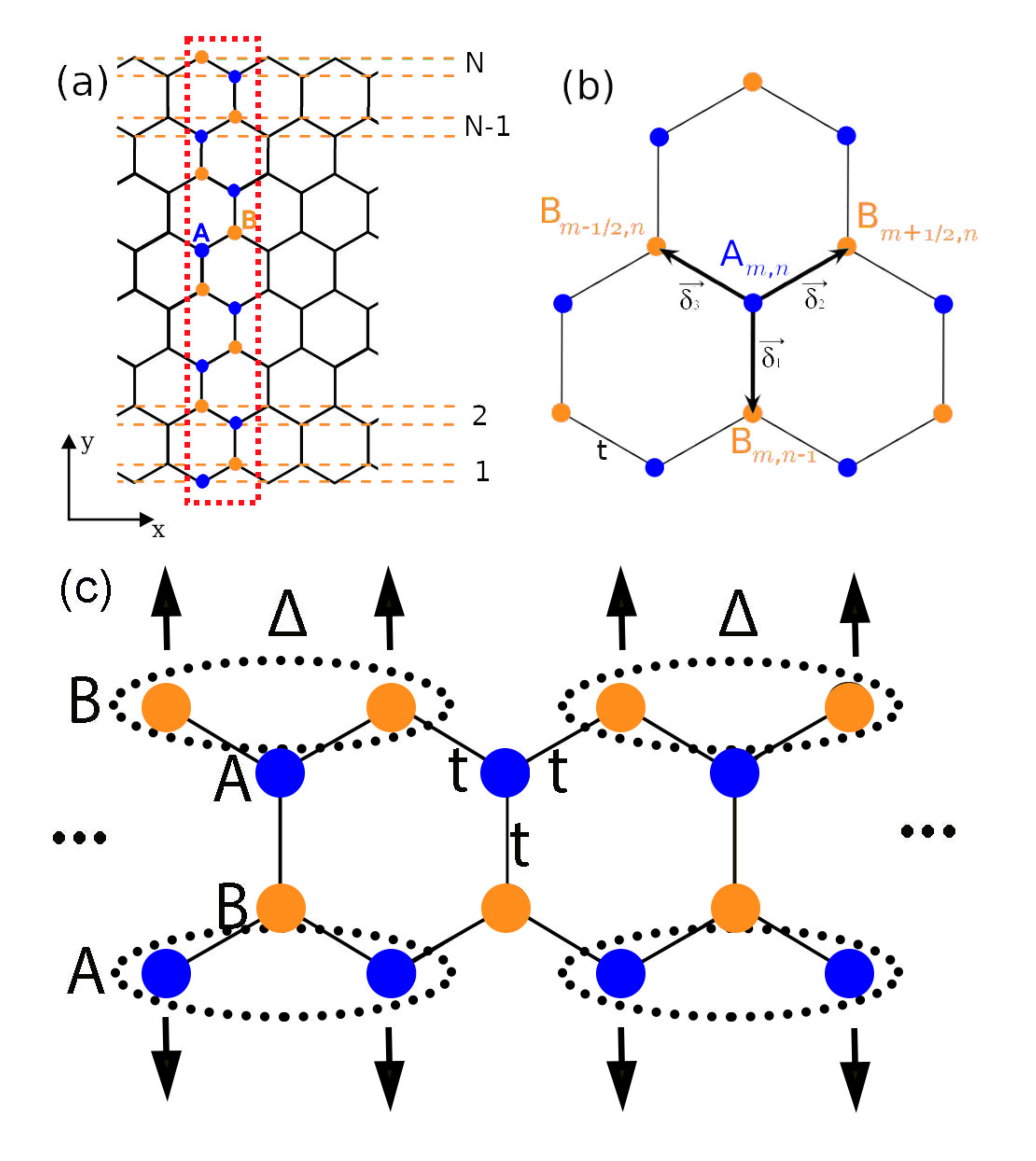}}\caption{(a) Sketch of the 2D zHNR geometry adopted here, where $N$ represents its width ($n=1,\cdots,N$). The region within the red dashed area composed of $2N$ nonequivalent $A$ (blue) and $B$ (orange) sites along the $y$ direction represents the unit cell employed in the numerical simulations. The $M$ number of unit cells defines the nanoribbon length ($m=1,\cdots,M$). (b) Representation of the nearest-neighbor hopping $t$,  which is adopted as the energy unit. (c) Schematic of a double-spin KzHNR of width $N=2$. The equivalent $B$ ($A$) atoms of the upper (lower) KzHNR are paired with each other via a \textit{p}-wave superconductivity parameter $\Delta$. 
	\label{fig:SystemGeometry}} 
\end{figure}

In this Letter, we report the possibility of spin discriminates MZMs in zHNRs geometry [Fig.~\ref{fig:SystemGeometry}(a)], which we refer to as double-spin Kitaev zigzag honeycomb nanoribbons (KzHNR). This double nanoribbon structure mimics two parallel Kitaev chains connected by the hopping $t$, as indicated in Fig.~\ref{fig:SystemGeometry}(c). Our findings reveal that we can assess the spin species of the MZMs in the double spin KzHNR by tuning the chemical potential of the chains. Moreover, we present an experimental proposal to discriminate spin-polarized MZMs in zHNRs structures of silicene that grows over a Pb superconductor in the presence of RSOC and an EMF. Our findings could contribute to paving the way for studying hybrid topological-conventional qubits using Majorana spintronics.


\textit{Spinless model and topological phase transitions---.} We first consider a double-spinless KzHNR as a generalization of the Kitaev chain~\cite{Kitaev2001} to characterize the topological phase transitions (TPT) of the system through corresponding winding numbers~\cite{Zhou2017}, computed for the infinite case (more details in the Note $1$ of SM). By considering a tight-binding chain in zHNR geometry, we define a spinless phenomenological model as a Kitaev ladder-type~\cite{Maiellaro2018}. We represent in Fig.~\ref{fig:SystemGeometry}(c), the first nearest neighbor (NN) hopping $t$ between nonequivalent sites $A$ and $B$ and the \textit{p}-superconducting wave pairing, indicated by arrows,  between equivalent sites $A$ or $B$ located at the edges of the KzHNR. The Hamiltonian describing such model reads

\begin{equation}
H = H_{t}+H_{\Delta}, \label{H_base}
\end{equation}
where, 
\begin{equation}
	\begin{split} 
		&H_{t}=\sum_{m,n}[ t ( a_{m,n}^{\dagger}b_{m,n-1} + a_{m,n}^{\dagger}b_{m-1/2,n}+\\
		& a_{m,n}^{\dagger}b_{m+1/2,n})
		- \sum_{n} \mu [a_{n,n}^{\dagger}a_{n,n} + b_{n,n}^{\dagger}b_{n,n}]+\text{H.c.}] ,
	\end{split}
	\label{Ht}
\end{equation}
corresponds to the NN hopping term $t$, as indicated in Fig.~\ref{fig:SystemGeometry}(b), where, $\mu$ is the chemical potential and the operators $a^{\dagger}_{m,n}/b_{m,n}$  creates/annihilates an electron at site  $A/B$ in the unit cell. Moreover, the Hamiltonian 
\begin{equation} 
\begin{split}
&H_{\Delta}  = \sum_{m,n} \Delta[a_{m,n}^{\dagger}a_{m+1,n}^{\dagger} - a_{m,n}^{\dagger}a_{m-1,n}^{\dagger}+ \\
&b_{m,n+1}^{\dagger}b_{m+1,n+1}^{\dagger} - b_{m,n+1}^{\dagger}b_{m-1,n+1}^{\dagger} +\text{H.c.}] , 
\end{split}
\label{Super}
\end{equation}
describes the \textit{p}-wave superconducting pairing of the double-spinless KzHNR, where $\Delta$ is the pairing strength between sites $B$ in the top and between sites $A$ in the bottom of each KzHNR, as indicated in Fig.~\ref{fig:SystemGeometry}(c). Once particle-hole, time-reversal, and chiral symmetries are preserved by the Hamiltonian [Eqs.~(\ref{H_base})-(\ref{Super})], it belongs to the BDI symmetry group class with $\mathbb{Z}$ index~\cite{Chiu2016,wakatsuki2014majorana} (see Note 1 of SM). For simplicity, we only have considered the double-spinless KzHNR of width $N=2$ in our numerical simulations. However, the results presented here are also valid for larger widths of nanoribbons. 
\begin{figure}[t]
	\centerline{\includegraphics[width=3.5in,keepaspectratio]{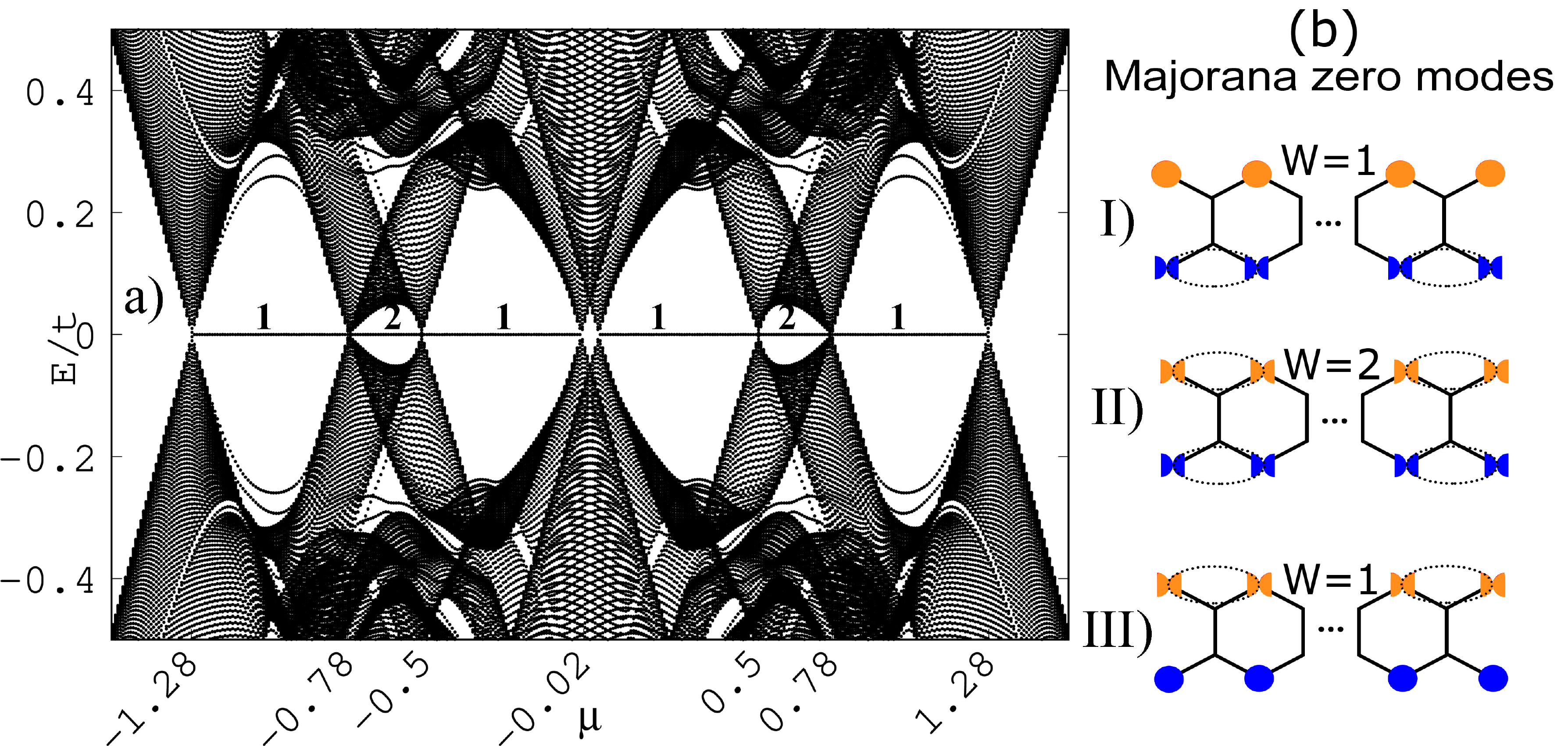}}\caption{(a) Energy spectrum of a 2D double-spinless KzHNR [Eq.~(\ref{H_base})] as a function of $\mu$ and \textit{p}-wave $\Delta = 0.5t$ for KzHNR of width $N=2$ and length $M=200$. The numbers on the real axis represent the $W$ associated with the corresponding topological region. (b) Schematic representation of the emergence of MZMs at the ends of the KzHNR for each associated $W$. Each semicircle represents an MZM generated on the site of the active border of the KzHNR. The two Majoranas connected with the dotted ellipses form a standard fermion. In the topological phase, unpaired Majorana fermions emerge at both ends of each KzHNR, as represented by the semicircles outside the dotted ellipses. The situations (I) and (III) describes $W=1$, where only either the top or bottom KzHNR is active to generate one MZM at each of its ends. Otherwise, in (II) $W=2$, indicating that both the KzHNRs generate MZMs simultaneously at their corresponding ends. 
	\label{fig:FiniteSpinlessKZHNR}}
\end{figure}

Fig.~\ref{fig:FiniteSpinlessKZHNR}(a) shows the bulk energy spectrum $E/t$ of the double-spinless KzHNR [Eq.~(\ref{H_base})] as a function of the chemical potential $\mu$. Several TPTs characterized by the closing-and-reopening of superconducting induced gap appear at the $\mu=-1.28t$, $-0.78t$, $-0.50t$, and $-0.02t$, respectively. According to the bulk-boundary correspondence principle~\cite{Alase2019}, the multiplicities of modes on the boundaries associated with the TPTs in bulk are characterized by topological invariants of the bulk energy bands, so-called winding number $W$, for instance. Here, we have found $W=0$, $W = 1$ and $W = 2$ [see Fig.~1(a)-(d) of the SM], which are indicated in the corresponding regions of Fig.~\ref{fig:FiniteSpinlessKZHNR}(a). Following the bulk-boundary principle, $W=0$ ($\mu>|1.28t|$) characterizes the trivial phase, where MZMs are absent at the ends of both the KzHNRs. Between $\mu = -1.28t$ and $\mu=-0.78t$, a topological region is characterized by $W=1$, indicating the emergence of MZMs at opposite ends of either bottom or top KzHNR [Fig.\ref{fig:FiniteSpinlessKZHNR}(b), cases (I) and (III)].

Fig.~\ref{fig:FiniteSpinlessKZHNR}(a) also exhibits another topological region characterized by $W=2$ in the interval $-0.78t<\mu<-0.50t$, for instance, corresponding to the situation in which the MZMs arises simultaneously in the ends of both top and bottom KzHNR, as indicated in the sketch (II) of Fig.~\ref{fig:FiniteSpinlessKZHNR}(b). In Fig.~\ref{fig:FiniteSpinlessKZHNR}(a), we also can notice that the same TPTs which occur for $\mu < 0$ appear for positive values of $\mu$ due to the particle-hole symmetry exhibited by the Hamiltonian of Eq.~(\ref{H_base})~\footnote{It also should be noticed in Fig.~\ref{fig:FiniteSpinlessKZHNR}(a) that there is an uncharacterized small region in the interval $-0.02t<\mu<0.02t$, which is an effect produced by the finite length of the KzHNR chain considered in the calculations ($M=200$) and therefore tends to disappear for larger values of $M$, giving rise to a single TPT at $\mu = 0$.}.

\textit{Spin full model and emergence of spin-polarized MZMs---.} To analyze the possibility of distinguishing the spin species of the MZMs, we now discuss the emergence of MZMs at the double-spin KzHNR geometry edges, considering both spin orientations explicitly. We account for the infinite version of the whole spin case in Note 1 of the SM (Fig. 1(e-k)). To properly break the spin degeneracy of the system, we introduce two additional effects in the Hamiltonian of Eq.~(\ref{H_base}): the extrinsic RSOC and an EMF. The extrinsic RSOC lifts the corresponding bands' spin degeneracy, unless at $\mathbf{k}=0$. Additionally, the EMF applied perpendicularly to the ribbon plane drives the system through TPTs exhibiting spin-polarized MZMs. In this situation, spin-discriminated MZMs emerge at the ends of the double-spin KzHNR structure. The corresponding generalized Hamiltonian is given by
\begin{equation}
 H=H_{t}+H_{\Delta} +H_{R}+H_{z},
\label{H_base1}   
\end{equation}
which can be written in a new basis of four distinct atoms, [as indicated in Fig. 2 of the Note 2 in the SM]. In this basis, the Hamiltonian describing the NN hopping and the \textit{p-}wave superconducting pairing reads
\begin{equation} \label{first1}
    \begin{split}
&H_{t} =  -\frac{\mu}{2}  \sum^M_{i,\sigma} (a_{i \sigma}^{\dagger}a_{i \sigma} - a_{i \sigma}a_{i \sigma}^{\dagger}+ b_{i \sigma}^{\dagger}b_{i \sigma} - b_{i \sigma}b_{i \sigma}^{\dagger}+\\
&c_{i \sigma}^{\dagger}c_{i \sigma} - c_{i \sigma}c_{i \sigma}^{\dagger}+d_{i \sigma}^{\dagger}d_{i \sigma} - d_{i \sigma}d_{i \sigma}^{\dagger})-\\ 
&\frac{t}{4} \sum^{M}_i (a_{i \sigma}^{\dagger}b_{i \sigma} - b_{i \sigma}a_{i \sigma}^{\dagger}+b_{i \sigma}^{\dagger}c_{i \sigma} - c_{i \sigma}b_{i \sigma}^{\dagger}+d_{i \sigma}^{\dagger}c_{i \sigma} - c_{i \sigma}d_{i \sigma}^{\dagger})- \\
 &\sum^{M-1}_i \frac{t}{4}(a_{i \sigma}^{\dagger}b_{i+1 \sigma} - b_{i+1 \sigma}a_{i \sigma}^{\dagger}+d_{i \sigma}^{\dagger}c_{i+1 \sigma} - c_{i+1 \sigma}d_{i \sigma}^{\dagger})+\text{H.c.},
    \end{split}
\end{equation}
\begin{equation} \label{first2}
    \begin{split}
&H_{\Delta}  =  \sum^{M-1}_i [\Delta (a_{i \sigma}a_{i+1 \sigma} -  a_{i+1 \sigma} a_{i \sigma} +\\
& d_{i \sigma}d_{i+1 \sigma} -  d_{i+1 \sigma} d_{i \sigma}) +\text{H.c.}].
    \end{split}
\end{equation}

The extrinsic RSOC induced in the KzHNR can be generated by breaking the inversion symmetry due to either a substrate with strong spin-orbit interaction~\cite{Perge14} or modulated by the action of an external electric field  $\vec{E}$ applied perpendicularly to the nanoribbon plane \cite{Min2006,zarea2009,Ezawa2012,Drummond2012,Latge2015,Jiao2020} and is given in general  by 
\begin{equation}
H_R= \sum_{i,j, \sigma}[i a_{i, \sigma}^{\dagger}(\vec{u}_{ij}.\vec{\sigma})a_{j, \sigma}+\text{H.c.}],
\label{Rashba}    
\end{equation}
where the $\vec{u}_{ij}$ is given by
$\vec{u}_{ij}= \left(\frac{e}{2m^{2}av_{f}}\right)\vec{E}\times\vec{\delta}_{ij}= -\frac{\lambda_{R}}{a}\hat{k}\times\vec{\delta}_{ij}$,
with $e$ and $m$ being the charge and mass of the electron respectively. Moreover, $v_{f}$ is the Fermi velocity, the lattice constant is given by $a$ and the vector-position $\vec{\delta}_{ij}$ corresponds to the three nearest neighbors, as represented in Fig.~1(b). Writting  Eq. \ref{Rashba} in the basis of Fig. 2 of  Note 2 in the SM, we can write the Rashba Hamiltonian as
\begin{equation}\label{MHR1}
\begin{split}
   &H_{R}=\sum^{M}_{i,\sigma} i \lambda_R sign(\sigma) \big[\gamma_{1}(a_{i\sigma}^{\dagger}b_{i\overline{\sigma}} - b_{i\sigma}a_{i\overline{\sigma}}^{\dagger}) + \\
   &\left(-\frac{1}{2}\right)(b_{i\sigma}^{\dagger}c_{i\overline{\sigma}} -c_{i\sigma}b_{i\overline{\sigma}}^{\dagger})+
  \gamma_{2} (c_{i\sigma}^{\dagger}d_{i\overline{\sigma}} - d_{i\sigma}c_{i\overline{\sigma}}^{\dagger})  \big]  \\
   & +\sum^{M-1}_{i,\sigma} i \lambda_R sign(\sigma)  \big[\gamma_{2}(b_{i\sigma}^{\dagger}a_{i+1\overline{\sigma}} - a_{i+1\sigma}b_{i\overline{\sigma}}^{\dagger})+ \\
   &\gamma_{1}(d_{i\sigma}^{\dagger}c_{i+1\overline{\sigma}} - c_{i+1\sigma}d_{i\overline{\sigma}}^{\dagger})  \big]  + \text{H.c.} ,
 \end{split}
\end{equation}
where $\gamma_{1}=\left(\frac{1}{2}+i\frac{\sqrt{3}}{2}\right)$, $\gamma_{2}=\left(\frac{1}{2}-i\frac{\sqrt{3}}{2}\right)$, $\lambda_{R}$ is the extrinsic RSOC parameter and $\sigma=\uparrow,\downarrow$ is the spin index for each operator. The last term of Eq.~(\ref{H_base1})
\begin{equation} \label{MHV2}
    \begin{split}
&H_{z}  = \frac{1}{2}   \sum^M_{i,\sigma} \lambda_{z} sign(\sigma)[(a_{i,\sigma}^{\dagger}a_{i,\sigma} - a_{i,\sigma}a_{i,\sigma} ^{\dagger})+ \\
&(b_{i,\sigma} ^{\dagger}b_{i,\sigma} - b_{i,\sigma}b_{i,\sigma} ^{\dagger})  
+(c_{i,\sigma} ^{\dagger}c_{i,\sigma} - c_ic_{i,\sigma}^{\dagger})+\\
&(d_{i,\sigma}^{\dagger}d_{i,\sigma} - d_{i,\sigma}d_{i,\sigma} ^{\dagger})]+\text{H.c.},  
    \end{split}
\end{equation}
represents an EMF with the magnetization vector pointing to the azimuthal direction ~\cite{ginetom, macdonald2011}, where $\lambda_{z}$ is the EMF strength. Details concerning the derivation of the extrinsic RSOC [Eq.~(\ref{MHR1})] and EMF [Eq.~(\ref{MHV2})] Hamiltonians are account in Note 2 of SM. In Fig.~3 of SM, we perform a detailed analysis of how the KzHNR length $M$, the extrinsic RSOC $\lambda_{R}$, the superconducting pairing $\Delta$ and the EMF $\lambda_{z}$ affects the emergence of MZMs on the real axis.

The emergence of MZMs polarized at the edges of the double-spin KzHNR is calculated by computing the mean value of $\braket{S_{z}}= \bra{\psi}\hat{S}_{z}\ket{\psi}$ of MZMs solutions. The label $\psi$ represents eigenvectors of the total Hamiltonian given by Eq.~(\ref{H_base1}) and $\hat{S}_{z}$ is the Pauli matrix in the $\hat{z}$ direction.

\begin{figure}[t]
\centerline{\includegraphics[width=3.5in,keepaspectratio]{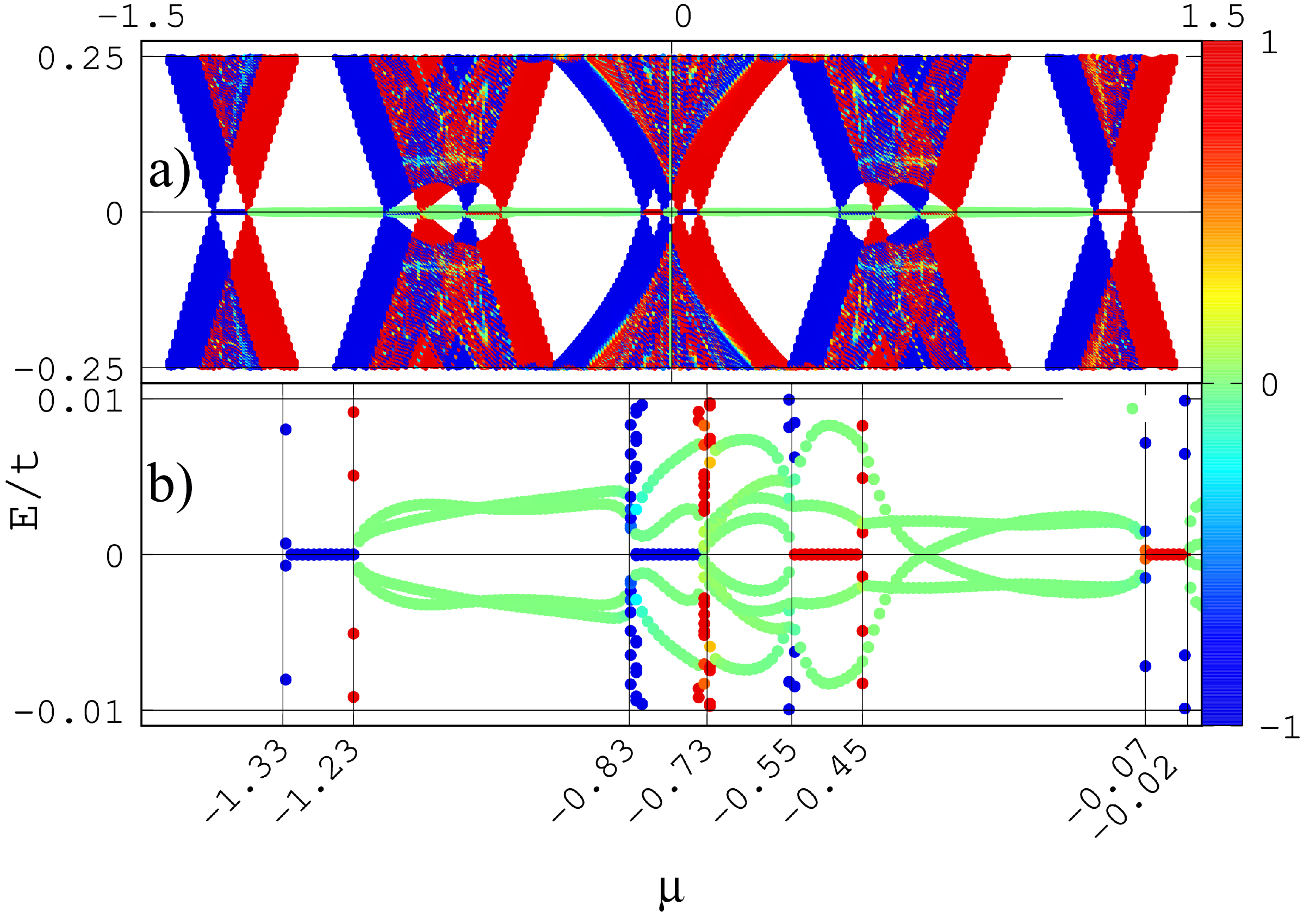}}\caption{(a)Energy spectra of the finite double-spin KzHNR [Eq.~\ref{H_base1}] as a function of $\mu$, $N=2$ and $M=200$. The blue and red colors correspond to spin up and down regions at the real axis, respectively. The green lines describe the formation of a regular fermion coming from the combination of Majorana excitations belonging to opposite KzHNRs. (b) Zoomed region of (a) around $E=0$ for $\mu<0$, showing in detail the formation of spin-polarized MZMs.
	\label{fig:FiniteFullCase}}
\end{figure}

Fig.~\ref{fig:FiniteFullCase}(a) shows a high-resolution energy spectrum $E/t$ of a finite double-spin KzHNR [Eq.~(\ref{H_base1})] with $N=2$ and $M=200$ as a function of $\mu$. It has the same shape as the double-spinless case [Fig.~\ref{fig:FiniteSpinlessKZHNR}(a)], but with spin-polarized energies resolved into spin-up (blue color) and spin-down (red color) regions at the real axis. A mirror spin-symmetry concerning $\mu =0$ is observed in the plot: a spin-up MZMs for $\mu<0$ changes to spin-down for $\mu>0$. Moreover, spin-polarized MZMs can be accessed by tunning $\mu$ slightly below or above $\mu=0$.

Fig.~\ref{fig:FiniteFullCase}(b) shows the zoomed region of (a) around $E=0$ for $\mu<0$, where it is possible to see in detail the emergence of spin-polarized MZMs as $\mu$ changes. We can detect these MZMs states with well-defined spin orientation via spin-polarized STM measurements~\cite{Jeon_2017}. The green lines depicted in both panels of Fig.~\ref{fig:FiniteFullCase} do not represent MZMs, but ordinary fermions, resulting from the combination of MZMs localized at the ends of opposite KzHNRs. This effect tends to disappear as the width $N$ of the double-spin KzHNR increases.

As discussed in the spinless case of Fig.~\ref{fig:FiniteSpinlessKZHNR}, the emergence of spin-polarized MZMs on the real axis depicted in Fig.~\ref{fig:FiniteFullCase} is also related to TPTs in the bulk gap. However, each value of $\mu$ related to a TPT in the spinless scenario splits into two values of $\mu$, describing TPTs for both spin up and down components. The strength of this splitting is given by the RSOC parameter $\lambda_{R}$. For details, see Fig.~1 and related discussion in the SM.      

 \textit{Experimental perspectives---.} Among available experimental results for realizing a double-spin KzHNR structure, we suggest the silicene deposited on a lead superconducting substrate~\cite{Krawiec2015,Krawiec2016,Krawiec2019,Owczarek2020} as a possible candidate. In the superconducting phase, under the presence of strong RSOC coming from the Pb and an applied EMF, the Cooper pairs of the Pb substrate can enter into the silicene region via proximity effect, giving rise to a \textit{p}-wave-induced pairing in the double KzHNR structure.

Moreover, several previous theoretical works have shown that the zHNRs accumulate electrons to form localized magnetic moments \cite{Nakada96} at its edges.  The coupling of atoms belonging to the same edge is ferromagnetic and between atoms from different edges is antiferromagnetic ~\cite{katsunori2010, Fu_2017, Jorge18, Zhu_2019}. This situation is depicted in Fig.~\ref{fig:SystemGeometry}(c). In particular, low-width silicene nanoribbons are predicted to have an antiferromagnetic ground state~\cite{Chengyong2012, Xiao_2018}. Another possibility to experimentally realize a double spin-polarized KzHNR is following the recipe of the reference~\cite{Perge14}: growing an antiferromagnetic nanoribbon or some artificial antiferromagnetic ladder over a strong spin-orbit conventional \textit{s}-wave superconductor.

\textit{ Conclusions---.}
This Letter reports the possibility of obtaining spin-polarized MZMs at opposite edges of a double-spin KzHNR structure. The regions of energy spectrum $E/t$ with MZMs having well-defined either spin up or down orientations can be accessed by tunning the $\mu$ of the KzHNRs. Moreover, these spin-polarized intervals in the $E/t \times \mu$ profile are associated with distinct topological phases, characterized by the topological invariant winding number $W=1$ or $W=2$. Interestingly enough, for the situation wherein $W=2$ four MZMs emerge in the double KzHNR geometry: two at the opposite ends of the top zHNR and two at the opposite ends of the bottom one. In this scenario, it should be emphasized that at least four MZMs are required for defining a qubit~\cite{KarzigPRB2017,NayakRMP2008,Steiner2020,AasenPRX2016}. Thus, the proposal is a natural candidate for realizing hybrid quantum computing operations~\cite{LeijnsePRL2011,Flensberg2012} between conventional spin qubits and topological qubits based on MZMs with well-defined spin orientation, suggesting a possible route for performing Majorana spintronics.

\textit{Acknowledgments---.} We acknowledge the support from Brazilian National Council for Scientific and Technological Development (CNPq) and Coordena\c{c}\~{a}o de Aperfei\c{c}oamento de Pessoal de N\'{i}vel Superior - Brasil (CAPES). M.~S.~F. Acknowledges financial support from the Brazilian agency Funda\c c\~ao de Amparo a Pesquisa do Estado do Rio de Janeiro, process 210 355/2018. L. S. R. acknowledges the Icelandic Research Fund (project ``Hybrid polaritonics''). A.C.S acknowledges CNPq Grant. Nr. 305668/2018-8.
\bibliography{references}

\begin{thebibliography}{62}%
\makeatletter
\providecommand \@ifxundefined [1]{%
 \@ifx{#1\undefined}
}%
\providecommand \@ifnum [1]{%
 \ifnum #1\expandafter \@firstoftwo
 \else \expandafter \@secondoftwo
 \fi
}%
\providecommand \@ifx [1]{%
 \ifx #1\expandafter \@firstoftwo
 \else \expandafter \@secondoftwo
 \fi
}%
\providecommand \natexlab [1]{#1}%
\providecommand \enquote  [1]{``#1''}%
\providecommand \bibnamefont  [1]{#1}%
\providecommand \bibfnamefont [1]{#1}%
\providecommand \citenamefont [1]{#1}%
\providecommand \href@noop [0]{\@secondoftwo}%
\providecommand \href [0]{\begingroup \@sanitize@url \@href}%
\providecommand \@href[1]{\@@startlink{#1}\@@href}%
\providecommand \@@href[1]{\endgroup#1\@@endlink}%
\providecommand \@sanitize@url [0]{\catcode `\\12\catcode `\$12\catcode
  `\&12\catcode `\#12\catcode `\^12\catcode `\_12\catcode `\%12\relax}%
\providecommand \@@startlink[1]{}%
\providecommand \@@endlink[0]{}%
\providecommand \url  [0]{\begingroup\@sanitize@url \@url }%
\providecommand \@url [1]{\endgroup\@href {#1}{\urlprefix }}%
\providecommand \urlprefix  [0]{URL }%
\providecommand \Eprint [0]{\href }%
\providecommand \doibase [0]{https://doi.org/}%
\providecommand \selectlanguage [0]{\@gobble}%
\providecommand \bibinfo  [0]{\@secondoftwo}%
\providecommand \bibfield  [0]{\@secondoftwo}%
\providecommand \translation [1]{[#1]}%
\providecommand \BibitemOpen [0]{}%
\providecommand \bibitemStop [0]{}%
\providecommand \bibitemNoStop [0]{.\EOS\space}%
\providecommand \EOS [0]{\spacefactor3000\relax}%
\providecommand \BibitemShut  [1]{\csname bibitem#1\endcsname}%
\let\auto@bib@innerbib\@empty
\bibitem [{\citenamefont {Read}\ and\ \citenamefont {Green}(2000)}]{Read20}%
  \BibitemOpen
  \bibfield  {author} {\bibinfo {author} {\bibfnamefont {N.}~\bibnamefont
  {Read}}\ and\ \bibinfo {author} {\bibfnamefont {D.}~\bibnamefont {Green}},\
  }\bibfield  {title} {\bibinfo {title} {Paired states of fermions in two
  dimensions with breaking of parity and time-reversal symmetries and the
  fractional quantum hall effect},\ }\href
  {https://doi.org/10.1103/PhysRevB.61.10267} {\bibfield  {journal} {\bibinfo
  {journal} {Phys. Rev. B}\ }\textbf {\bibinfo {volume} {61}},\ \bibinfo
  {pages} {10267} (\bibinfo {year} {2000})}\BibitemShut {NoStop}%
\bibitem [{\citenamefont {Kitaev}(2001)}]{Kitaev2001}%
  \BibitemOpen
  \bibfield  {author} {\bibinfo {author} {\bibfnamefont {A.~Y.}\ \bibnamefont
  {Kitaev}},\ }\bibfield  {title} {\bibinfo {title} {Unpaired majorana fermions
  in quantum wires},\ }\href {http://stacks.iop.org/1063-7869/44/i=10S/a=S29}
  {\bibfield  {journal} {\bibinfo  {journal} {Physics-Uspekhi}\ }\textbf
  {\bibinfo {volume} {44}},\ \bibinfo {pages} {131} (\bibinfo {year}
  {2001})}\BibitemShut {NoStop}%
\bibitem [{\citenamefont {Oreg}\ \emph {et~al.}(2010)\citenamefont {Oreg},
  \citenamefont {Refael},\ and\ \citenamefont {von Oppen}}]{Oreg10}%
  \BibitemOpen
  \bibfield  {author} {\bibinfo {author} {\bibfnamefont {Y.}~\bibnamefont
  {Oreg}}, \bibinfo {author} {\bibfnamefont {G.}~\bibnamefont {Refael}},\ and\
  \bibinfo {author} {\bibfnamefont {F.}~\bibnamefont {von Oppen}},\ }\bibfield
  {title} {\bibinfo {title} {Helical liquids and majorana bound states in
  quantum wires},\ }\href {https://doi.org/10.1103/PhysRevLett.105.177002}
  {\bibfield  {journal} {\bibinfo  {journal} {Phys. Rev. Lett.}\ }\textbf
  {\bibinfo {volume} {105}},\ \bibinfo {pages} {177002} (\bibinfo {year}
  {2010})}\BibitemShut {NoStop}%
\bibitem [{\citenamefont {Lutchyn}\ \emph {et~al.}(2010)\citenamefont
  {Lutchyn}, \citenamefont {Sau},\ and\ \citenamefont {Das~Sarma}}]{Lutchin10}%
  \BibitemOpen
  \bibfield  {author} {\bibinfo {author} {\bibfnamefont {R.~M.}\ \bibnamefont
  {Lutchyn}}, \bibinfo {author} {\bibfnamefont {J.~D.}\ \bibnamefont {Sau}},\
  and\ \bibinfo {author} {\bibfnamefont {S.}~\bibnamefont {Das~Sarma}},\
  }\bibfield  {title} {\bibinfo {title} {Majorana fermions and a topological
  phase transition in semiconductor-superconductor heterostructures},\ }\href
  {https://doi.org/10.1103/PhysRevLett.105.077001} {\bibfield  {journal}
  {\bibinfo  {journal} {Phys. Rev. Lett.}\ }\textbf {\bibinfo {volume} {105}},\
  \bibinfo {pages} {077001} (\bibinfo {year} {2010})}\BibitemShut {NoStop}%
\bibitem [{\citenamefont {Mourik}\ \emph {et~al.}(2012)\citenamefont {Mourik},
  \citenamefont {Zuo}, \citenamefont {Frolov}, \citenamefont {Plissard},
  \citenamefont {Bakkers},\ and\ \citenamefont {Kouwenhoven}}]{Mourik12}%
  \BibitemOpen
  \bibfield  {author} {\bibinfo {author} {\bibfnamefont {V.}~\bibnamefont
  {Mourik}}, \bibinfo {author} {\bibfnamefont {K.}~\bibnamefont {Zuo}},
  \bibinfo {author} {\bibfnamefont {S.~M.}\ \bibnamefont {Frolov}}, \bibinfo
  {author} {\bibfnamefont {S.~R.}\ \bibnamefont {Plissard}}, \bibinfo {author}
  {\bibfnamefont {E.~P. A.~M.}\ \bibnamefont {Bakkers}},\ and\ \bibinfo
  {author} {\bibfnamefont {L.~P.}\ \bibnamefont {Kouwenhoven}},\ }\bibfield
  {title} {\bibinfo {title} {Signatures of majorana fermions in hybrid
  superconductor-semiconductor nanowire devices},\ }\href
  {https://doi.org/10.1126/science.1222360} {\bibfield  {journal} {\bibinfo
  {journal} {Science}\ }\textbf {\bibinfo {volume} {336}},\ \bibinfo {pages}
  {1003} (\bibinfo {year} {2012})}\BibitemShut {NoStop}%
\bibitem [{\citenamefont {G{\"u}l}\ \emph {et~al.}(2018)\citenamefont
  {G{\"u}l}, \citenamefont {Zhang}, \citenamefont {Bommer}, \citenamefont
  {de~Moor}, \citenamefont {Car}, \citenamefont {Plissard}, \citenamefont
  {Bakkers}, \citenamefont {Geresdi}, \citenamefont {Watanabe}, \citenamefont
  {Taniguchi},\ and\ \citenamefont {Kouwenhoven}}]{Gul2018}%
  \BibitemOpen
  \bibfield  {author} {\bibinfo {author} {\bibfnamefont {{\"O}.}~\bibnamefont
  {G{\"u}l}}, \bibinfo {author} {\bibfnamefont {H.}~\bibnamefont {Zhang}},
  \bibinfo {author} {\bibfnamefont {J.~D.~S.}\ \bibnamefont {Bommer}}, \bibinfo
  {author} {\bibfnamefont {M.~W.~A.}\ \bibnamefont {de~Moor}}, \bibinfo
  {author} {\bibfnamefont {D.}~\bibnamefont {Car}}, \bibinfo {author}
  {\bibfnamefont {S.~R.}\ \bibnamefont {Plissard}}, \bibinfo {author}
  {\bibfnamefont {E.~P. A.~M.}\ \bibnamefont {Bakkers}}, \bibinfo {author}
  {\bibfnamefont {A.}~\bibnamefont {Geresdi}}, \bibinfo {author} {\bibfnamefont
  {K.}~\bibnamefont {Watanabe}}, \bibinfo {author} {\bibfnamefont
  {T.}~\bibnamefont {Taniguchi}},\ and\ \bibinfo {author} {\bibfnamefont
  {L.~P.}\ \bibnamefont {Kouwenhoven}},\ }\bibfield  {title} {\bibinfo {title}
  {Ballistic majorana nanowire devices},\ }\href
  {https://doi.org/10.1038/s41565-017-0032-8} {\bibfield  {journal} {\bibinfo
  {journal} {Nature Nanotechnology}\ }\textbf {\bibinfo {volume} {13}},\
  \bibinfo {pages} {192} (\bibinfo {year} {2018})}\BibitemShut {NoStop}%
\bibitem [{\citenamefont {Aguado}(2017)}]{Aguado17}%
  \BibitemOpen
  \bibfield  {author} {\bibinfo {author} {\bibfnamefont {R.}~\bibnamefont
  {Aguado}},\ }\bibfield  {title} {\bibinfo {title} {Majorana quasiparticles in
  condensed matter},\ }\href {https://doi.org/10.1393/ncr/i2017-10141-9}
  {\bibfield  {journal} {\bibinfo  {journal} {Riv Nuovo Cimento}\ }\textbf
  {\bibinfo {volume} {40}},\ \bibinfo {pages} {523} (\bibinfo {year}
  {2017})}\BibitemShut {NoStop}%
\bibitem [{\citenamefont {Krogstrup}\ \emph {et~al.}(2015)\citenamefont
  {Krogstrup}, \citenamefont {Ziino}, \citenamefont {Chang}, \citenamefont
  {Albrecht}, \citenamefont {Madsen}, \citenamefont {Johnson}, \citenamefont
  {Nyg{\aa}rd}, \citenamefont {Marcus},\ and\ \citenamefont
  {Jespersen}}]{Krogstrup2015}%
  \BibitemOpen
  \bibfield  {author} {\bibinfo {author} {\bibfnamefont {P.}~\bibnamefont
  {Krogstrup}}, \bibinfo {author} {\bibfnamefont {N.~L.~B.}\ \bibnamefont
  {Ziino}}, \bibinfo {author} {\bibfnamefont {W.}~\bibnamefont {Chang}},
  \bibinfo {author} {\bibfnamefont {S.~M.}\ \bibnamefont {Albrecht}}, \bibinfo
  {author} {\bibfnamefont {M.~H.}\ \bibnamefont {Madsen}}, \bibinfo {author}
  {\bibfnamefont {E.}~\bibnamefont {Johnson}}, \bibinfo {author} {\bibfnamefont
  {J.}~\bibnamefont {Nyg{\aa}rd}}, \bibinfo {author} {\bibfnamefont {C.~M.}\
  \bibnamefont {Marcus}},\ and\ \bibinfo {author} {\bibfnamefont {T.~S.}\
  \bibnamefont {Jespersen}},\ }\bibfield  {title} {\bibinfo {title} {Epitaxy of
  semiconductor--superconductor nanowires},\ }\href
  {https://doi.org/10.1038/nmat4176} {\bibfield  {journal} {\bibinfo  {journal}
  {Nature Materials}\ }\textbf {\bibinfo {volume} {14}},\ \bibinfo {pages}
  {400} (\bibinfo {year} {2015})}\BibitemShut {NoStop}%
\bibitem [{\citenamefont {Clarke}(2017)}]{Clarke2017}%
  \BibitemOpen
  \bibfield  {author} {\bibinfo {author} {\bibfnamefont {D.~J.}\ \bibnamefont
  {Clarke}},\ }\bibfield  {title} {\bibinfo {title} {Experimentally accessible
  topological quality factor for wires with zero energy modes},\ }\href
  {https://doi.org/10.1103/PhysRevB.96.201109} {\bibfield  {journal} {\bibinfo
  {journal} {Phys. Rev. B}\ }\textbf {\bibinfo {volume} {96}},\ \bibinfo
  {pages} {201109} (\bibinfo {year} {2017})}\BibitemShut {NoStop}%
\bibitem [{\citenamefont {Schuray}\ \emph {et~al.}(2017)\citenamefont
  {Schuray}, \citenamefont {Weithofer},\ and\ \citenamefont
  {Recher}}]{Schuray2017}%
  \BibitemOpen
  \bibfield  {author} {\bibinfo {author} {\bibfnamefont {A.}~\bibnamefont
  {Schuray}}, \bibinfo {author} {\bibfnamefont {L.}~\bibnamefont {Weithofer}},\
  and\ \bibinfo {author} {\bibfnamefont {P.}~\bibnamefont {Recher}},\
  }\bibfield  {title} {\bibinfo {title} {Fano resonances in majorana bound
  states--quantum dot hybrid systems},\ }\href
  {https://doi.org/10.1103/PhysRevB.96.085417} {\bibfield  {journal} {\bibinfo
  {journal} {Phys. Rev. B}\ }\textbf {\bibinfo {volume} {96}},\ \bibinfo
  {pages} {085417} (\bibinfo {year} {2017})}\BibitemShut {NoStop}%
\bibitem [{\citenamefont {Prada}\ \emph {et~al.}(2017)\citenamefont {Prada},
  \citenamefont {Aguado},\ and\ \citenamefont {San-Jose}}]{Prada2018}%
  \BibitemOpen
  \bibfield  {author} {\bibinfo {author} {\bibfnamefont {E.}~\bibnamefont
  {Prada}}, \bibinfo {author} {\bibfnamefont {R.}~\bibnamefont {Aguado}},\ and\
  \bibinfo {author} {\bibfnamefont {P.}~\bibnamefont {San-Jose}},\ }\bibfield
  {title} {\bibinfo {title} {Measuring majorana nonlocality and spin structure
  with a quantum dot},\ }\href {https://doi.org/10.1103/PhysRevB.96.085418}
  {\bibfield  {journal} {\bibinfo  {journal} {Phys. Rev. B}\ }\textbf {\bibinfo
  {volume} {96}},\ \bibinfo {pages} {085418} (\bibinfo {year}
  {2017})}\BibitemShut {NoStop}%
\bibitem [{\citenamefont {Ricco}\ \emph {et~al.}(2019)\citenamefont {Ricco},
  \citenamefont {de~Souza}, \citenamefont {Figueira}, \citenamefont {Shelykh},\
  and\ \citenamefont {Seridonio}}]{Ricco2018}%
  \BibitemOpen
  \bibfield  {author} {\bibinfo {author} {\bibfnamefont {L.~S.}\ \bibnamefont
  {Ricco}}, \bibinfo {author} {\bibfnamefont {M.}~\bibnamefont {de~Souza}},
  \bibinfo {author} {\bibfnamefont {M.~S.}\ \bibnamefont {Figueira}}, \bibinfo
  {author} {\bibfnamefont {I.~A.}\ \bibnamefont {Shelykh}},\ and\ \bibinfo
  {author} {\bibfnamefont {A.~C.}\ \bibnamefont {Seridonio}},\ }\bibfield
  {title} {\bibinfo {title} {Spin-dependent zero-bias peak in a hybrid
  nanowire-quantum dot system: Distinguishing isolated majorana fermions from
  andreev bound states},\ }\href {https://doi.org/10.1103/PhysRevB.99.155159}
  {\bibfield  {journal} {\bibinfo  {journal} {Phys. Rev. B}\ }\textbf {\bibinfo
  {volume} {99}},\ \bibinfo {pages} {155159} (\bibinfo {year}
  {2019})}\BibitemShut {NoStop}%
\bibitem [{\citenamefont {Deng}\ \emph {et~al.}(2016)\citenamefont {Deng},
  \citenamefont {Vaitiekenas}, \citenamefont {Hansen}, \citenamefont {Danon},
  \citenamefont {Leijnse}, \citenamefont {Flensberg}, \citenamefont {Nyg{\r
  a}rd}, \citenamefont {Krogstrup},\ and\ \citenamefont {Marcus}}]{Deng2016}%
  \BibitemOpen
  \bibfield  {author} {\bibinfo {author} {\bibfnamefont {M.~T.}\ \bibnamefont
  {Deng}}, \bibinfo {author} {\bibfnamefont {S.}~\bibnamefont {Vaitiekenas}},
  \bibinfo {author} {\bibfnamefont {E.~B.}\ \bibnamefont {Hansen}}, \bibinfo
  {author} {\bibfnamefont {J.}~\bibnamefont {Danon}}, \bibinfo {author}
  {\bibfnamefont {M.}~\bibnamefont {Leijnse}}, \bibinfo {author} {\bibfnamefont
  {K.}~\bibnamefont {Flensberg}}, \bibinfo {author} {\bibfnamefont
  {J.}~\bibnamefont {Nyg{\r a}rd}}, \bibinfo {author} {\bibfnamefont
  {P.}~\bibnamefont {Krogstrup}},\ and\ \bibinfo {author} {\bibfnamefont
  {C.~M.}\ \bibnamefont {Marcus}},\ }\bibfield  {title} {\bibinfo {title}
  {Majorana bound state in a coupled quantum-dot hybrid-nanowire system},\
  }\href {https://doi.org/10.1126/science.aaf3961} {\bibfield  {journal}
  {\bibinfo  {journal} {Science}\ }\textbf {\bibinfo {volume} {354}},\ \bibinfo
  {pages} {1557} (\bibinfo {year} {2016})}\BibitemShut {NoStop}%
\bibitem [{\citenamefont {Deng}\ \emph {et~al.}(2018)\citenamefont {Deng},
  \citenamefont {Vaitiek\ifmmode~\dot{e}\else \.{e}\fi{}nas}, \citenamefont
  {Prada}, \citenamefont {San-Jose}, \citenamefont {Nyg\aa{}rd}, \citenamefont
  {Krogstrup}, \citenamefont {Aguado},\ and\ \citenamefont
  {Marcus}}]{Deng2018}%
  \BibitemOpen
  \bibfield  {author} {\bibinfo {author} {\bibfnamefont {M.-T.}\ \bibnamefont
  {Deng}}, \bibinfo {author} {\bibfnamefont {S.}~\bibnamefont
  {Vaitiek\ifmmode~\dot{e}\else \.{e}\fi{}nas}}, \bibinfo {author}
  {\bibfnamefont {E.}~\bibnamefont {Prada}}, \bibinfo {author} {\bibfnamefont
  {P.}~\bibnamefont {San-Jose}}, \bibinfo {author} {\bibfnamefont
  {J.}~\bibnamefont {Nyg\aa{}rd}}, \bibinfo {author} {\bibfnamefont
  {P.}~\bibnamefont {Krogstrup}}, \bibinfo {author} {\bibfnamefont
  {R.}~\bibnamefont {Aguado}},\ and\ \bibinfo {author} {\bibfnamefont {C.~M.}\
  \bibnamefont {Marcus}},\ }\bibfield  {title} {\bibinfo {title} {Nonlocality
  of majorana modes in hybrid nanowires},\ }\href
  {https://doi.org/10.1103/PhysRevB.98.085125} {\bibfield  {journal} {\bibinfo
  {journal} {Phys. Rev. B}\ }\textbf {\bibinfo {volume} {98}},\ \bibinfo
  {pages} {085125} (\bibinfo {year} {2018})}\BibitemShut {NoStop}%
\bibitem [{\citenamefont {Nadj-Perge}\ \emph {et~al.}(2014)\citenamefont
  {Nadj-Perge}, \citenamefont {Drozdov}, \citenamefont {Li}, \citenamefont
  {Chen}, \citenamefont {Jeon}, \citenamefont {Seo}, \citenamefont {MacDonald},
  \citenamefont {Bernevig},\ and\ \citenamefont {Yazdani}}]{Perge14}%
  \BibitemOpen
  \bibfield  {author} {\bibinfo {author} {\bibfnamefont {S.}~\bibnamefont
  {Nadj-Perge}}, \bibinfo {author} {\bibfnamefont {I.~K.}\ \bibnamefont
  {Drozdov}}, \bibinfo {author} {\bibfnamefont {J.}~\bibnamefont {Li}},
  \bibinfo {author} {\bibfnamefont {H.}~\bibnamefont {Chen}}, \bibinfo {author}
  {\bibfnamefont {S.}~\bibnamefont {Jeon}}, \bibinfo {author} {\bibfnamefont
  {J.}~\bibnamefont {Seo}}, \bibinfo {author} {\bibfnamefont {A.~H.}\
  \bibnamefont {MacDonald}}, \bibinfo {author} {\bibfnamefont {B.~A.}\
  \bibnamefont {Bernevig}},\ and\ \bibinfo {author} {\bibfnamefont
  {A.}~\bibnamefont {Yazdani}},\ }\bibfield  {title} {\bibinfo {title}
  {Observation of majorana fermions in ferromagnetic atomic chains on a
  superconductor},\ }\href {https://doi.org/10.1126/science.1259327} {\bibfield
   {journal} {\bibinfo  {journal} {Science}\ }\textbf {\bibinfo {volume}
  {346}},\ \bibinfo {pages} {602} (\bibinfo {year} {2014})}\BibitemShut
  {NoStop}%
\bibitem [{\citenamefont {Jeon}\ \emph
  {et~al.}(2017{\natexlab{a}})\citenamefont {Jeon}, \citenamefont {Xie},
  \citenamefont {Li}, \citenamefont {Wang}, \citenamefont {Bernevig},\ and\
  \citenamefont {Yazdani}}]{JeonScience2017}%
  \BibitemOpen
  \bibfield  {author} {\bibinfo {author} {\bibfnamefont {S.}~\bibnamefont
  {Jeon}}, \bibinfo {author} {\bibfnamefont {Y.}~\bibnamefont {Xie}}, \bibinfo
  {author} {\bibfnamefont {J.}~\bibnamefont {Li}}, \bibinfo {author}
  {\bibfnamefont {Z.}~\bibnamefont {Wang}}, \bibinfo {author} {\bibfnamefont
  {B.~A.}\ \bibnamefont {Bernevig}},\ and\ \bibinfo {author} {\bibfnamefont
  {A.}~\bibnamefont {Yazdani}},\ }\bibfield  {title} {\bibinfo {title}
  {Distinguishing a majorana zero mode using spin-resolved measurements},\
  }\href {https://doi.org/10.1126/science.aan3670} {\bibfield  {journal}
  {\bibinfo  {journal} {Science}\ }\textbf {\bibinfo {volume} {358}},\ \bibinfo
  {pages} {772} (\bibinfo {year} {2017}{\natexlab{a}})}\BibitemShut {NoStop}%
\bibitem [{\citenamefont {Flensberg}\ \emph {et~al.}(2021)\citenamefont
  {Flensberg}, \citenamefont {von Oppen},\ and\ \citenamefont
  {Stern}}]{Flensberg2021}%
  \BibitemOpen
  \bibfield  {author} {\bibinfo {author} {\bibfnamefont {K.}~\bibnamefont
  {Flensberg}}, \bibinfo {author} {\bibfnamefont {F.}~\bibnamefont {von
  Oppen}},\ and\ \bibinfo {author} {\bibfnamefont {A.}~\bibnamefont {Stern}},\
  }\bibfield  {title} {\bibinfo {title} {Engineered platforms for topological
  superconductivity and majorana zero modes},\ }\bibfield  {journal} {\bibinfo
  {journal} {Nature Reviews Materials}\ }\href
  {https://doi.org/10.1038/s41578-021-00336-6} {10.1038/s41578-021-00336-6}
  (\bibinfo {year} {2021})\BibitemShut {NoStop}%
\bibitem [{\citenamefont {Berthold}\ \emph {et~al.}(2021)\citenamefont
  {Berthold}, \citenamefont {Yonglong},\ and\ \citenamefont
  {Yazdani}}]{NatureReviewBerthold2021}%
  \BibitemOpen
  \bibfield  {author} {\bibinfo {author} {\bibfnamefont {J.}~\bibnamefont
  {Berthold}}, \bibinfo {author} {\bibfnamefont {X.}~\bibnamefont {Yonglong}},\
  and\ \bibinfo {author} {\bibfnamefont {A.}~\bibnamefont {Yazdani}},\
  }\bibfield  {title} {\bibinfo {title} {Detecting and distinguishing majorana
  zero modes with the scanning tunnelling microscope},\ }\bibfield  {journal}
  {\bibinfo  {journal} {Nat Rev Phys}\ }\href
  {https://doi.org/10.1038/s42254-021-00328-z} {10.1038/s42254-021-00328-z}
  (\bibinfo {year} {2021})\BibitemShut {NoStop}%
\bibitem [{\citenamefont {Zhang}\ \emph {et~al.}(2019)\citenamefont {Zhang},
  \citenamefont {Liu}, \citenamefont {Wimmer},\ and\ \citenamefont
  {Kouwenhoven}}]{ZhangNatCommun2019}%
  \BibitemOpen
  \bibfield  {author} {\bibinfo {author} {\bibfnamefont {H.}~\bibnamefont
  {Zhang}}, \bibinfo {author} {\bibfnamefont {D.~E.}\ \bibnamefont {Liu}},
  \bibinfo {author} {\bibfnamefont {M.}~\bibnamefont {Wimmer}},\ and\ \bibinfo
  {author} {\bibfnamefont {L.~P.}\ \bibnamefont {Kouwenhoven}},\ }\bibfield
  {title} {\bibinfo {title} {Next steps of quantum transport in majorana
  nanowire devices},\ }\href {https://doi.org/10.1038/s41467-019-13133-1}
  {\bibfield  {journal} {\bibinfo  {journal} {Nat Commun}\ }\textbf {\bibinfo
  {volume} {10}},\ \bibinfo {pages} {5128} (\bibinfo {year}
  {2019})}\BibitemShut {NoStop}%
\bibitem [{\citenamefont {Prada}\ \emph {et~al.}(2020)\citenamefont {Prada},
  \citenamefont {San-Jose}, \citenamefont {de~Moor}, \citenamefont {Geresdi},
  \citenamefont {Lee}, \citenamefont {Klinovaja}, \citenamefont {Loss},
  \citenamefont {Nygård}, \citenamefont {Aguado},\ and\ \citenamefont
  {Kouwenhoven}}]{prada2020andreev}%
  \BibitemOpen
  \bibfield  {author} {\bibinfo {author} {\bibfnamefont {E.}~\bibnamefont
  {Prada}}, \bibinfo {author} {\bibfnamefont {P.}~\bibnamefont {San-Jose}},
  \bibinfo {author} {\bibfnamefont {M.~W.~A.}\ \bibnamefont {de~Moor}},
  \bibinfo {author} {\bibfnamefont {A.}~\bibnamefont {Geresdi}}, \bibinfo
  {author} {\bibfnamefont {E.~J.~H.}\ \bibnamefont {Lee}}, \bibinfo {author}
  {\bibfnamefont {J.}~\bibnamefont {Klinovaja}}, \bibinfo {author}
  {\bibfnamefont {D.}~\bibnamefont {Loss}}, \bibinfo {author} {\bibfnamefont
  {J.}~\bibnamefont {Nygård}}, \bibinfo {author} {\bibfnamefont
  {R.}~\bibnamefont {Aguado}},\ and\ \bibinfo {author} {\bibfnamefont {L.~P.}\
  \bibnamefont {Kouwenhoven}},\ }\bibfield  {title} {\bibinfo {title} {From
  andreev to majorana bound states in hybrid superconductor-semiconductor
  nanowires},\ }\href {https://doi.org/10.1038/s42254-020-0228-y} {\bibfield
  {journal} {\bibinfo  {journal} {Nat Rev Phys}\ }\textbf {\bibinfo {volume}
  {2}},\ \bibinfo {pages} {575} (\bibinfo {year} {2020})}\BibitemShut {NoStop}%
\bibitem [{\citenamefont {Zhaoa}\ and\ \citenamefont {Wang}(2020)}]{Aidi20}%
  \BibitemOpen
  \bibfield  {author} {\bibinfo {author} {\bibfnamefont {A.}~\bibnamefont
  {Zhaoa}}\ and\ \bibinfo {author} {\bibfnamefont {B.}~\bibnamefont {Wang}},\
  }\bibfield  {title} {\bibinfo {title} {Two-dimensional graphene-like xenes as
  potential topological materials},\ }\href {https://doi.org/10.1063/1.5135984}
  {\bibfield  {journal} {\bibinfo  {journal} {APL Materials}\ }\textbf
  {\bibinfo {volume} {8}},\ \bibinfo {pages} {030701} (\bibinfo {year}
  {2020})}\BibitemShut {NoStop}%
\bibitem [{\citenamefont {Dutreix}\ \emph {et~al.}(2014)\citenamefont
  {Dutreix}, \citenamefont {Guigou}, \citenamefont {Chevallier},\ and\
  \citenamefont {Bena}}]{Dutreix2014}%
  \BibitemOpen
  \bibfield  {author} {\bibinfo {author} {\bibfnamefont {C.}~\bibnamefont
  {Dutreix}}, \bibinfo {author} {\bibfnamefont {M.}~\bibnamefont {Guigou}},
  \bibinfo {author} {\bibfnamefont {D.}~\bibnamefont {Chevallier}},\ and\
  \bibinfo {author} {\bibfnamefont {C.}~\bibnamefont {Bena}},\ }\bibfield
  {title} {\bibinfo {title} {Majorana fermions in honeycomb lattices},\ }\href
  {https://doi.org/10.1140/epjb/e2014-50243-9} {\bibfield  {journal} {\bibinfo
  {journal} {The European Physical Journal B}\ }\textbf {\bibinfo {volume}
  {87}},\ \bibinfo {pages} {296} (\bibinfo {year} {2014})}\BibitemShut
  {NoStop}%
\bibitem [{\citenamefont {Ma}\ \emph {et~al.}(2017)\citenamefont {Ma},
  \citenamefont {Yang}, \citenamefont {Huang},\ and\ \citenamefont
  {Lin}}]{Ma2017}%
  \BibitemOpen
  \bibfield  {author} {\bibinfo {author} {\bibfnamefont {T.}~\bibnamefont
  {Ma}}, \bibinfo {author} {\bibfnamefont {F.}~\bibnamefont {Yang}}, \bibinfo
  {author} {\bibfnamefont {Z.}~\bibnamefont {Huang}},\ and\ \bibinfo {author}
  {\bibfnamefont {H.-Q.}\ \bibnamefont {Lin}},\ }\bibfield  {title} {\bibinfo
  {title} {Triplet p-wave pairing correlation in low-doped zigzag graphene
  nanoribbons},\ }\href {https://doi.org/10.1038/srep42262} {\bibfield
  {journal} {\bibinfo  {journal} {Scientific Reports}\ }\textbf {\bibinfo
  {volume} {7}},\ \bibinfo {pages} {42262 EP } (\bibinfo {year} {2017})},\
  \bibinfo {note} {article}\BibitemShut {NoStop}%
\bibitem [{\citenamefont {Lalmi}\ \emph {et~al.}(2010)\citenamefont {Lalmi},
  \citenamefont {Oughaddou}, \citenamefont {Enriquez}, \citenamefont {Kara},
  \citenamefont {Vizzini}, \citenamefont {Ealet},\ and\ \citenamefont
  {Aufray}}]{Lalmi2010}%
  \BibitemOpen
  \bibfield  {author} {\bibinfo {author} {\bibfnamefont {B.}~\bibnamefont
  {Lalmi}}, \bibinfo {author} {\bibfnamefont {H.}~\bibnamefont {Oughaddou}},
  \bibinfo {author} {\bibfnamefont {H.}~\bibnamefont {Enriquez}}, \bibinfo
  {author} {\bibfnamefont {A.}~\bibnamefont {Kara}}, \bibinfo {author}
  {\bibfnamefont {S.}~\bibnamefont {Vizzini}}, \bibinfo {author} {\bibfnamefont
  {B.}~\bibnamefont {Ealet}},\ and\ \bibinfo {author} {\bibfnamefont
  {B.}~\bibnamefont {Aufray}},\ }\bibfield  {title} {\bibinfo {title}
  {Epitaxial growth of a silicene sheet},\ }\href
  {https://doi.org/10.1063/1.3524215} {\bibfield  {journal} {\bibinfo
  {journal} {Applied Physics Letters}\ }\textbf {\bibinfo {volume} {97}},\
  \bibinfo {pages} {223109} (\bibinfo {year} {2010})}\BibitemShut {NoStop}%
\bibitem [{\citenamefont {Aufray}\ \emph {et~al.}(2010)\citenamefont {Aufray},
  \citenamefont {Kara}, \citenamefont {Vizzini}, \citenamefont {Oughaddou},
  \citenamefont {L\'eandri}, \citenamefont {Ealet},\ and\ \citenamefont
  {Le~Lay}}]{Aufray2010}%
  \BibitemOpen
  \bibfield  {author} {\bibinfo {author} {\bibfnamefont {B.}~\bibnamefont
  {Aufray}}, \bibinfo {author} {\bibfnamefont {A.}~\bibnamefont {Kara}},
  \bibinfo {author} {\bibfnamefont {S.}~\bibnamefont {Vizzini}}, \bibinfo
  {author} {\bibfnamefont {H.}~\bibnamefont {Oughaddou}}, \bibinfo {author}
  {\bibfnamefont {C.}~\bibnamefont {L\'eandri}}, \bibinfo {author}
  {\bibfnamefont {B.}~\bibnamefont {Ealet}},\ and\ \bibinfo {author}
  {\bibfnamefont {G.}~\bibnamefont {Le~Lay}},\ }\bibfield  {title} {\bibinfo
  {title} {Graphene-like silicon nanoribbons on ag(110): A possible formation
  of silicene},\ }\href@noop {} {\bibfield  {journal} {\bibinfo  {journal}
  {Applied Physics Letters}\ }\textbf {\bibinfo {volume} {96}},\ \bibinfo
  {pages} {183102} (\bibinfo {year} {2010})}\BibitemShut {NoStop}%
\bibitem [{\citenamefont {Liu}\ \emph {et~al.}(2011)\citenamefont {Liu},
  \citenamefont {Jiang},\ and\ \citenamefont {Yao}}]{Liu2011}%
  \BibitemOpen
  \bibfield  {author} {\bibinfo {author} {\bibfnamefont {C.-C.}\ \bibnamefont
  {Liu}}, \bibinfo {author} {\bibfnamefont {H.}~\bibnamefont {Jiang}},\ and\
  \bibinfo {author} {\bibfnamefont {Y.}~\bibnamefont {Yao}},\ }\bibfield
  {title} {\bibinfo {title} {Low-energy effective hamiltonian involving
  spin-orbit coupling in silicene and two-dimensional germanium and tin},\
  }\href {https://doi.org/10.1103/PhysRevB.84.195430} {\bibfield  {journal}
  {\bibinfo  {journal} {Phys. Rev. B}\ }\textbf {\bibinfo {volume} {84}},\
  \bibinfo {pages} {195430} (\bibinfo {year} {2011})}\BibitemShut {NoStop}%
\bibitem [{\citenamefont {Ezawa}(2012)}]{Ezawa2012}%
  \BibitemOpen
  \bibfield  {author} {\bibinfo {author} {\bibfnamefont {M.}~\bibnamefont
  {Ezawa}},\ }\bibfield  {title} {\bibinfo {title} {Valley-polarized metals and
  quantum anomalous hall effect in silicene},\ }\href
  {https://doi.org/10.1103/PhysRevLett.109.055502} {\bibfield  {journal}
  {\bibinfo  {journal} {Phys. Rev. Lett.}\ }\textbf {\bibinfo {volume} {109}},\
  \bibinfo {pages} {055502} (\bibinfo {year} {2012})}\BibitemShut {NoStop}%
\bibitem [{\citenamefont {Drummond}\ \emph {et~al.}(2012)\citenamefont
  {Drummond}, \citenamefont {Z\'olyomi},\ and\ \citenamefont
  {Fal'ko}}]{Drummond2012}%
  \BibitemOpen
  \bibfield  {author} {\bibinfo {author} {\bibfnamefont {N.~D.}\ \bibnamefont
  {Drummond}}, \bibinfo {author} {\bibfnamefont {V.}~\bibnamefont
  {Z\'olyomi}},\ and\ \bibinfo {author} {\bibfnamefont {V.~I.}\ \bibnamefont
  {Fal'ko}},\ }\bibfield  {title} {\bibinfo {title} {Electrically tunable band
  gap in silicene},\ }\href {https://doi.org/10.1103/PhysRevB.85.075423}
  {\bibfield  {journal} {\bibinfo  {journal} {Phys. Rev. B}\ }\textbf {\bibinfo
  {volume} {85}},\ \bibinfo {pages} {075423} (\bibinfo {year}
  {2012})}\BibitemShut {NoStop}%
\bibitem [{\citenamefont {Jiao}\ \emph {et~al.}(2020)\citenamefont {Jiao},
  \citenamefont {Yao},\ and\ \citenamefont {Zandvliet}}]{Jiao2020}%
  \BibitemOpen
  \bibfield  {author} {\bibinfo {author} {\bibfnamefont {Z.}~\bibnamefont
  {Jiao}}, \bibinfo {author} {\bibfnamefont {Q.}~\bibnamefont {Yao}},\ and\
  \bibinfo {author} {\bibfnamefont {H.}~\bibnamefont {Zandvliet}},\ }\bibfield
  {title} {\bibinfo {title} {Tailoring and probing the quantum states of matter
  of $2d$ dirac materials with a buckled honeycomb structure},\ }\href
  {https://doi.org/10.1016/j.physe.2020.114113} {\bibfield  {journal} {\bibinfo
   {journal} {Physica E: Low-dimensional Systems and Nanostructures}\ }\textbf
  {\bibinfo {volume} {121}},\ \bibinfo {pages} {114113} (\bibinfo {year}
  {2020})}\BibitemShut {NoStop}%
\bibitem [{\citenamefont {Tsai}\ \emph {et~al.}(2013)\citenamefont {Tsai},
  \citenamefont {Huang}, \citenamefont {Chang}, \citenamefont {Lin},
  \citenamefont {Jeng},\ and\ \citenamefont {Bansil}}]{Tsai2013}%
  \BibitemOpen
  \bibfield  {author} {\bibinfo {author} {\bibfnamefont {W.-F.}\ \bibnamefont
  {Tsai}}, \bibinfo {author} {\bibfnamefont {C.-Y.}\ \bibnamefont {Huang}},
  \bibinfo {author} {\bibfnamefont {T.-R.}\ \bibnamefont {Chang}}, \bibinfo
  {author} {\bibfnamefont {H.}~\bibnamefont {Lin}}, \bibinfo {author}
  {\bibfnamefont {H.-T.}\ \bibnamefont {Jeng}},\ and\ \bibinfo {author}
  {\bibfnamefont {A.}~\bibnamefont {Bansil}},\ }\bibfield  {title} {\bibinfo
  {title} {Gated silicene as a tunable source of nearly 100{\%} spin-polarized
  electrons},\ }\href {https://doi.org/10.1038/ncomms2525} {\bibfield
  {journal} {\bibinfo  {journal} {Nature Communications}\ }\textbf {\bibinfo
  {volume} {4}},\ \bibinfo {pages} {1500} (\bibinfo {year} {2013})}\BibitemShut
  {NoStop}%
\bibitem [{\citenamefont {Chico}\ \emph {et~al.}(2015)\citenamefont {Chico},
  \citenamefont {Latgé},\ and\ \citenamefont {Brey}}]{Latge2015}%
  \BibitemOpen
  \bibfield  {author} {\bibinfo {author} {\bibfnamefont {L.}~\bibnamefont
  {Chico}}, \bibinfo {author} {\bibfnamefont {A.}~\bibnamefont {Latgé}},\ and\
  \bibinfo {author} {\bibfnamefont {L.}~\bibnamefont {Brey}},\ }\bibfield
  {title} {\bibinfo {title} {Symmetries of quantum transport with rashba
  spin–orbit: graphene spintronics},\ }\href
  {https://doi.org/10.1039/C5CP01637A} {\bibfield  {journal} {\bibinfo
  {journal} {Phys. Chem. Chem. Phys.}\ }\textbf {\bibinfo {volume} {17}},\
  \bibinfo {pages} {16469} (\bibinfo {year} {2015})}\BibitemShut {NoStop}%
\bibitem [{\citenamefont {Jiang}\ \emph {et~al.}(2019)\citenamefont {Jiang},
  \citenamefont {Kang}, \citenamefont {Tao}, \citenamefont {Cao}, \citenamefont
  {Hao}, \citenamefont {Zheng}, \citenamefont {Zhang},\ and\ \citenamefont
  {Zeng}}]{Jiang_2019}%
  \BibitemOpen
  \bibfield  {author} {\bibinfo {author} {\bibfnamefont {P.}~\bibnamefont
  {Jiang}}, \bibinfo {author} {\bibfnamefont {L.}~\bibnamefont {Kang}},
  \bibinfo {author} {\bibfnamefont {X.}~\bibnamefont {Tao}}, \bibinfo {author}
  {\bibfnamefont {N.}~\bibnamefont {Cao}}, \bibinfo {author} {\bibfnamefont
  {H.}~\bibnamefont {Hao}}, \bibinfo {author} {\bibfnamefont {X.}~\bibnamefont
  {Zheng}}, \bibinfo {author} {\bibfnamefont {L.}~\bibnamefont {Zhang}},\ and\
  \bibinfo {author} {\bibfnamefont {Z.}~\bibnamefont {Zeng}},\ }\bibfield
  {title} {\bibinfo {title} {Robust generation of half-metallic transport and
  pure spin current with photogalvanic effect in zigzag silicene nanoribbons},\
  }\href {https://doi.org/10.1088/1361-648x/ab3dd6} {\bibfield  {journal}
  {\bibinfo  {journal} {Journal of Physics: Condensed Matter}\ }\textbf
  {\bibinfo {volume} {31}},\ \bibinfo {pages} {495701} (\bibinfo {year}
  {2019})}\BibitemShut {NoStop}%
\bibitem [{\citenamefont {Silva}\ \emph {et~al.}(2020)\citenamefont {Silva},
  \citenamefont {da~Silva},\ and\ \citenamefont {Vernek}}]{Vernek2020}%
  \BibitemOpen
  \bibfield  {author} {\bibinfo {author} {\bibfnamefont {J.~F.}\ \bibnamefont
  {Silva}}, \bibinfo {author} {\bibfnamefont {L.~G. G. V.~D.}\ \bibnamefont
  {da~Silva}},\ and\ \bibinfo {author} {\bibfnamefont {E.}~\bibnamefont
  {Vernek}},\ }\bibfield  {title} {\bibinfo {title} {Robustness of the kondo
  effect in a quantum dot coupled to majorana zero modes},\ }\href
  {https://doi.org/10.1103/PhysRevB.101.075428} {\bibfield  {journal} {\bibinfo
   {journal} {Phys. Rev. B}\ }\textbf {\bibinfo {volume} {101}},\ \bibinfo
  {pages} {075428} (\bibinfo {year} {2020})}\BibitemShut {NoStop}%
\bibitem [{\citenamefont {Schuray}\ \emph {et~al.}(2020)\citenamefont
  {Schuray}, \citenamefont {Rammler},\ and\ \citenamefont
  {Recher}}]{Schuray2020}%
  \BibitemOpen
  \bibfield  {author} {\bibinfo {author} {\bibfnamefont {A.}~\bibnamefont
  {Schuray}}, \bibinfo {author} {\bibfnamefont {M.}~\bibnamefont {Rammler}},\
  and\ \bibinfo {author} {\bibfnamefont {P.}~\bibnamefont {Recher}},\
  }\bibfield  {title} {\bibinfo {title} {Signatures of the majorana spin in
  electrical transport through a majorana nanowire},\ }\href
  {https://doi.org/10.1103/PhysRevB.102.045303} {\bibfield  {journal} {\bibinfo
   {journal} {Phys. Rev. B}\ }\textbf {\bibinfo {volume} {102}},\ \bibinfo
  {pages} {045303} (\bibinfo {year} {2020})}\BibitemShut {NoStop}%
\bibitem [{\citenamefont {Leijnse}\ and\ \citenamefont
  {Flensberg}(2011)}]{LeijnsePRL2011}%
  \BibitemOpen
  \bibfield  {author} {\bibinfo {author} {\bibfnamefont {M.}~\bibnamefont
  {Leijnse}}\ and\ \bibinfo {author} {\bibfnamefont {K.}~\bibnamefont
  {Flensberg}},\ }\bibfield  {title} {\bibinfo {title} {Quantum information
  transfer between topological and spin qubit systems},\ }\href
  {https://doi.org/10.1103/PhysRevLett.107.210502} {\bibfield  {journal}
  {\bibinfo  {journal} {Phys. Rev. Lett.}\ }\textbf {\bibinfo {volume} {107}},\
  \bibinfo {pages} {210502} (\bibinfo {year} {2011})}\BibitemShut {NoStop}%
\bibitem [{\citenamefont {Leijnse}\ and\ \citenamefont
  {Flensberg}(2012)}]{Flensberg2012}%
  \BibitemOpen
  \bibfield  {author} {\bibinfo {author} {\bibfnamefont {M.}~\bibnamefont
  {Leijnse}}\ and\ \bibinfo {author} {\bibfnamefont {K.}~\bibnamefont
  {Flensberg}},\ }\bibfield  {title} {\bibinfo {title} {Hybrid topological-spin
  qubit systems for two-qubit-spin gates},\ }\href
  {https://doi.org/10.1103/PhysRevB.86.104511} {\bibfield  {journal} {\bibinfo
  {journal} {Phys. Rev. B}\ }\textbf {\bibinfo {volume} {86}},\ \bibinfo
  {pages} {104511} (\bibinfo {year} {2012})}\BibitemShut {NoStop}%
\bibitem [{\citenamefont {Zhou}\ \emph {et~al.}(2017)\citenamefont {Zhou},
  \citenamefont {Xu},\ and\ \citenamefont {Zhou}}]{Zhou2017}%
  \BibitemOpen
  \bibfield  {author} {\bibinfo {author} {\bibfnamefont {B.-Z.}\ \bibnamefont
  {Zhou}}, \bibinfo {author} {\bibfnamefont {D.-H.}\ \bibnamefont {Xu}},\ and\
  \bibinfo {author} {\bibfnamefont {B.}~\bibnamefont {Zhou}},\ }\bibfield
  {title} {\bibinfo {title} {Majorana zero modes in a ladder of
  density-modulated kitaev superconductor chains},\ }\href
  {https://doi.org/https://doi.org/10.1016/j.physleta.2017.05.035} {\bibfield
  {journal} {\bibinfo  {journal} {Physics Letters A}\ }\textbf {\bibinfo
  {volume} {381}},\ \bibinfo {pages} {2426 } (\bibinfo {year}
  {2017})}\BibitemShut {NoStop}%
\bibitem [{\citenamefont {Maiellaro}\ \emph {et~al.}(2018)\citenamefont
  {Maiellaro}, \citenamefont {Romeo},\ and\ \citenamefont
  {Citro}}]{Maiellaro2018}%
  \BibitemOpen
  \bibfield  {author} {\bibinfo {author} {\bibfnamefont {A.}~\bibnamefont
  {Maiellaro}}, \bibinfo {author} {\bibfnamefont {F.}~\bibnamefont {Romeo}},\
  and\ \bibinfo {author} {\bibfnamefont {R.}~\bibnamefont {Citro}},\ }\bibfield
   {title} {\bibinfo {title} {Topological phase diagram of a kitaev ladder},\
  }\href {https://doi.org/10.1140/epjst/e2018-800090-y} {\bibfield  {journal}
  {\bibinfo  {journal} {The European Physical Journal Special Topics}\ }\textbf
  {\bibinfo {volume} {227}},\ \bibinfo {pages} {1397} (\bibinfo {year}
  {2018})}\BibitemShut {NoStop}%
\bibitem [{\citenamefont {Chiu}\ \emph {et~al.}(2016)\citenamefont {Chiu},
  \citenamefont {Teo}, \citenamefont {Schnyder},\ and\ \citenamefont
  {Ryu}}]{Chiu2016}%
  \BibitemOpen
  \bibfield  {author} {\bibinfo {author} {\bibfnamefont {C.-K.}\ \bibnamefont
  {Chiu}}, \bibinfo {author} {\bibfnamefont {J.~C.~Y.}\ \bibnamefont {Teo}},
  \bibinfo {author} {\bibfnamefont {A.~P.}\ \bibnamefont {Schnyder}},\ and\
  \bibinfo {author} {\bibfnamefont {S.}~\bibnamefont {Ryu}},\ }\bibfield
  {title} {\bibinfo {title} {Classification of topological quantum matter with
  symmetries},\ }\href {https://doi.org/10.1103/RevModPhys.88.035005}
  {\bibfield  {journal} {\bibinfo  {journal} {Rev. Mod. Phys.}\ }\textbf
  {\bibinfo {volume} {88}},\ \bibinfo {pages} {035005} (\bibinfo {year}
  {2016})}\BibitemShut {NoStop}%
\bibitem [{\citenamefont {Wakatsuki}\ \emph {et~al.}(2014)\citenamefont
  {Wakatsuki}, \citenamefont {Ezawa},\ and\ \citenamefont
  {Nagaosa}}]{wakatsuki2014majorana}%
  \BibitemOpen
  \bibfield  {author} {\bibinfo {author} {\bibfnamefont {R.}~\bibnamefont
  {Wakatsuki}}, \bibinfo {author} {\bibfnamefont {M.}~\bibnamefont {Ezawa}},\
  and\ \bibinfo {author} {\bibfnamefont {N.}~\bibnamefont {Nagaosa}},\
  }\bibfield  {title} {\bibinfo {title} {Majorana fermions and multiple
  topological phase transition in kitaev ladder topological superconductors},\
  }\href@noop {} {\bibfield  {journal} {\bibinfo  {journal} {Physical Review
  B}\ }\textbf {\bibinfo {volume} {89}},\ \bibinfo {pages} {174514} (\bibinfo
  {year} {2014})}\BibitemShut {NoStop}%
\bibitem [{\citenamefont {Alase}(2019)}]{Alase2019}%
  \BibitemOpen
  \bibfield  {author} {\bibinfo {author} {\bibfnamefont {A.}~\bibnamefont
  {Alase}},\ }\href@noop {} {\emph {\bibinfo {title} {Boundary Physics and
  Bulk-Boundary Correspondence in Topological Phases of Matter}}},\ \bibinfo
  {edition} {1st}\ ed.\ (\bibinfo  {publisher} {Springer Theses},\ \bibinfo
  {address} {Springer Nature Switzerland AG},\ \bibinfo {year}
  {2019})\BibitemShut {NoStop}%
\bibitem [{Note1()}]{Note1}%
  \BibitemOpen
  \bibinfo {note} {It also should be noticed in Fig.~\ref
  {fig:FiniteSpinlessKZHNR}(a) that there is an uncharacterized small region in
  the interval $-0.02t<\mu <0.02t$, which is an effect produced by the finite
  length of the KzHNR chain considered in the calculations ($M=200$) and
  therefore tends to disappear for larger values of $M$, giving rise to a
  single TPT at $\mu = 0$.}\BibitemShut {Stop}%
\bibitem [{\citenamefont {Min}\ \emph {et~al.}(2006)\citenamefont {Min},
  \citenamefont {Hill}, \citenamefont {Sinitsyn}, \citenamefont {Sahu},
  \citenamefont {Kleinman},\ and\ \citenamefont {MacDonald}}]{Min2006}%
  \BibitemOpen
  \bibfield  {author} {\bibinfo {author} {\bibfnamefont {H.}~\bibnamefont
  {Min}}, \bibinfo {author} {\bibfnamefont {J.~E.}\ \bibnamefont {Hill}},
  \bibinfo {author} {\bibfnamefont {N.~A.}\ \bibnamefont {Sinitsyn}}, \bibinfo
  {author} {\bibfnamefont {B.~R.}\ \bibnamefont {Sahu}}, \bibinfo {author}
  {\bibfnamefont {L.}~\bibnamefont {Kleinman}},\ and\ \bibinfo {author}
  {\bibfnamefont {A.~H.}\ \bibnamefont {MacDonald}},\ }\bibfield  {title}
  {\bibinfo {title} {Intrinsic and rashba spin-orbit interactions in graphene
  sheets},\ }\href {https://doi.org/10.1103/PhysRevB.74.165310} {\bibfield
  {journal} {\bibinfo  {journal} {Phys. Rev. B}\ }\textbf {\bibinfo {volume}
  {74}},\ \bibinfo {pages} {165310} (\bibinfo {year} {2006})}\BibitemShut
  {NoStop}%
\bibitem [{\citenamefont {Zarea}\ and\ \citenamefont
  {Sandler}(2009)}]{zarea2009}%
  \BibitemOpen
  \bibfield  {author} {\bibinfo {author} {\bibfnamefont {M.}~\bibnamefont
  {Zarea}}\ and\ \bibinfo {author} {\bibfnamefont {N.}~\bibnamefont
  {Sandler}},\ }\bibfield  {title} {\bibinfo {title} {Rashba spin-orbit
  interaction in graphene and zigzag nanoribbons},\ }\href
  {https://doi.org/10.1103/PhysRevB.79.165442} {\bibfield  {journal} {\bibinfo
  {journal} {Phys. Rev. B}\ }\textbf {\bibinfo {volume} {79}},\ \bibinfo
  {pages} {165442} (\bibinfo {year} {2009})}\BibitemShut {NoStop}%
\bibitem [{\citenamefont {Diniz}\ \emph {et~al.}(2014)\citenamefont {Diniz},
  \citenamefont {Guassi},\ and\ \citenamefont {Qu}}]{ginetom}%
  \BibitemOpen
  \bibfield  {author} {\bibinfo {author} {\bibfnamefont {G.~S.}\ \bibnamefont
  {Diniz}}, \bibinfo {author} {\bibfnamefont {M.~R.}\ \bibnamefont {Guassi}},\
  and\ \bibinfo {author} {\bibfnamefont {F.}~\bibnamefont {Qu}},\ }\bibfield
  {title} {\bibinfo {title} {Controllable spin-charge transport in strained
  graphene nanoribbon devices},\ }\href {https://doi.org/10.1063/1.4896251}
  {\bibfield  {journal} {\bibinfo  {journal} {Journal of Applied Physics}\
  }\textbf {\bibinfo {volume} {116}},\ \bibinfo {pages} {113705} (\bibinfo
  {year} {2014})}\BibitemShut {NoStop}%
\bibitem [{\citenamefont {Tse}\ \emph {et~al.}(2011)\citenamefont {Tse},
  \citenamefont {Qiao}, \citenamefont {Yao}, \citenamefont {MacDonald},\ and\
  \citenamefont {Niu}}]{macdonald2011}%
  \BibitemOpen
  \bibfield  {author} {\bibinfo {author} {\bibfnamefont {W.-K.}\ \bibnamefont
  {Tse}}, \bibinfo {author} {\bibfnamefont {Z.}~\bibnamefont {Qiao}}, \bibinfo
  {author} {\bibfnamefont {Y.}~\bibnamefont {Yao}}, \bibinfo {author}
  {\bibfnamefont {A.~H.}\ \bibnamefont {MacDonald}},\ and\ \bibinfo {author}
  {\bibfnamefont {Q.}~\bibnamefont {Niu}},\ }\bibfield  {title} {\bibinfo
  {title} {Quantum anomalous hall effect in single-layer and bilayer
  graphene},\ }\href {https://doi.org/10.1103/PhysRevB.83.155447} {\bibfield
  {journal} {\bibinfo  {journal} {Phys. Rev. B}\ }\textbf {\bibinfo {volume}
  {83}},\ \bibinfo {pages} {155447} (\bibinfo {year} {2011})}\BibitemShut
  {NoStop}%
\bibitem [{\citenamefont {Jeon}\ \emph
  {et~al.}(2017{\natexlab{b}})\citenamefont {Jeon}, \citenamefont {Xie},
  \citenamefont {Li}, \citenamefont {Wang}, \citenamefont {Bernevig},\ and\
  \citenamefont {Yazdani}}]{Jeon_2017}%
  \BibitemOpen
  \bibfield  {author} {\bibinfo {author} {\bibfnamefont {S.}~\bibnamefont
  {Jeon}}, \bibinfo {author} {\bibfnamefont {Y.}~\bibnamefont {Xie}}, \bibinfo
  {author} {\bibfnamefont {J.}~\bibnamefont {Li}}, \bibinfo {author}
  {\bibfnamefont {Z.}~\bibnamefont {Wang}}, \bibinfo {author} {\bibfnamefont
  {B.~A.}\ \bibnamefont {Bernevig}},\ and\ \bibinfo {author} {\bibfnamefont
  {A.}~\bibnamefont {Yazdani}},\ }\bibfield  {title} {\bibinfo {title}
  {Distinguishing a majorana zero mode using spin-resolved measurements},\
  }\href {https://doi.org/10.1126/science.aan3670} {\bibfield  {journal}
  {\bibinfo  {journal} {Science}\ }\textbf {\bibinfo {volume} {358}},\ \bibinfo
  {pages} {772} (\bibinfo {year} {2017}{\natexlab{b}})}\BibitemShut {NoStop}%
\bibitem [{\citenamefont {Podsiad\l{}y-Paszkowska}\ and\ \citenamefont
  {Krawiec}(2015)}]{Krawiec2015}%
  \BibitemOpen
  \bibfield  {author} {\bibinfo {author} {\bibfnamefont {A.}~\bibnamefont
  {Podsiad\l{}y-Paszkowska}}\ and\ \bibinfo {author} {\bibfnamefont
  {M.}~\bibnamefont {Krawiec}},\ }\bibfield  {title} {\bibinfo {title}
  {Silicene on metallic quantum wells: An efficient way of tuning
  silicene-substrate interaction},\ }\href
  {https://doi.org/10.1103/PhysRevB.92.165411} {\bibfield  {journal} {\bibinfo
  {journal} {Phys. Rev. B}\ }\textbf {\bibinfo {volume} {92}},\ \bibinfo
  {pages} {165411} (\bibinfo {year} {2015})}\BibitemShut {NoStop}%
\bibitem [{\citenamefont {Stepniak-Dybala}\ and\ \citenamefont
  {Krawiec}(2016)}]{Krawiec2016}%
  \BibitemOpen
  \bibfield  {author} {\bibinfo {author} {\bibfnamefont {M.}~\bibnamefont
  {Stepniak-Dybala}, \bibfnamefont {Jalochowski}}\ and\ \bibinfo {author}
  {\bibfnamefont {M.}~\bibnamefont {Krawiec}},\ }\bibfield  {title} {\bibinfo
  {title} {Silicene nanoribbons on pb-reconstructed si(111) surface},\
  }\bibfield  {journal} {\bibinfo  {journal} {Condens. Matter}\ }\textbf
  {\bibinfo {volume} {1}},\ \href {https://doi.org/10.3390/condmat1010008}
  {10.3390/condmat1010008} (\bibinfo {year} {2016})\BibitemShut {NoStop}%
\bibitem [{\citenamefont {Stepniak-Dybala}\ and\ \citenamefont
  {Krawiec}(2019)}]{Krawiec2019}%
  \BibitemOpen
  \bibfield  {author} {\bibinfo {author} {\bibfnamefont {A.}~\bibnamefont
  {Stepniak-Dybala}}\ and\ \bibinfo {author} {\bibfnamefont {M.}~\bibnamefont
  {Krawiec}},\ }\bibfield  {title} {\bibinfo {title} {Formation of silicene on
  ultrathin pb(111) films},\ }\href {https://doi.org/10.1021/acs.jpcc.9b04343}
  {\bibfield  {journal} {\bibinfo  {journal} {The Journal of Physical Chemistry
  C}\ }\textbf {\bibinfo {volume} {123}},\ \bibinfo {pages} {17019} (\bibinfo
  {year} {2019})}\BibitemShut {NoStop}%
\bibitem [{\citenamefont {Owczarek}\ and\ \citenamefont
  {Markowski}(2020)}]{Owczarek2020}%
  \BibitemOpen
  \bibfield  {author} {\bibinfo {author} {\bibfnamefont {S.}~\bibnamefont
  {Owczarek}}\ and\ \bibinfo {author} {\bibfnamefont {L.}~\bibnamefont
  {Markowski}},\ }\bibfield  {title} {\bibinfo {title} {The role of surfactant
  in two-components structures formation on si(111) surface},\ }\href
  {https://doi.org/https://doi.org/10.1016/j.susc.2019.121552} {\bibfield
  {journal} {\bibinfo  {journal} {Surface Science}\ }\textbf {\bibinfo {volume}
  {693}},\ \bibinfo {pages} {121552} (\bibinfo {year} {2020})}\BibitemShut
  {NoStop}%
\bibitem [{\citenamefont {Nakada}\ \emph {et~al.}(1996)\citenamefont {Nakada},
  \citenamefont {Fujita}, \citenamefont {Dresselhaus},\ and\ \citenamefont
  {Dresselhaus}}]{Nakada96}%
  \BibitemOpen
  \bibfield  {author} {\bibinfo {author} {\bibfnamefont {K.}~\bibnamefont
  {Nakada}}, \bibinfo {author} {\bibfnamefont {M.}~\bibnamefont {Fujita}},
  \bibinfo {author} {\bibfnamefont {G.}~\bibnamefont {Dresselhaus}},\ and\
  \bibinfo {author} {\bibfnamefont {M.~S.}\ \bibnamefont {Dresselhaus}},\
  }\bibfield  {title} {\bibinfo {title} {Edge state in graphene ribbons:
  Nanometer size effect and edge shape dependence},\ }\href
  {https://doi.org/10.1103/PhysRevB.54.17954} {\bibfield  {journal} {\bibinfo
  {journal} {Phys. Rev. B}\ }\textbf {\bibinfo {volume} {54}},\ \bibinfo
  {pages} {17954} (\bibinfo {year} {1996})}\BibitemShut {NoStop}%
\bibitem [{\citenamefont {Wakabayashi}\ \emph {et~al.}(2010)\citenamefont
  {Wakabayashi}, \citenamefont {ichi Sasaki}, \citenamefont {Nakanishi},\ and\
  \citenamefont {Enoki}}]{katsunori2010}%
  \BibitemOpen
  \bibfield  {author} {\bibinfo {author} {\bibfnamefont {K.}~\bibnamefont
  {Wakabayashi}}, \bibinfo {author} {\bibfnamefont {K.}~\bibnamefont {ichi
  Sasaki}}, \bibinfo {author} {\bibfnamefont {T.}~\bibnamefont {Nakanishi}},\
  and\ \bibinfo {author} {\bibfnamefont {T.}~\bibnamefont {Enoki}},\ }\bibfield
   {title} {\bibinfo {title} {Electronic states of graphene nanoribbons and
  analytical solutions},\ }\href
  {https://doi.org/10.1088/1468-6996/11/5/054504} {\bibfield  {journal}
  {\bibinfo  {journal} {Science and Technology of Advanced Materials}\ }\textbf
  {\bibinfo {volume} {11}},\ \bibinfo {pages} {054504} (\bibinfo {year}
  {2010})}\BibitemShut {NoStop}%
\bibitem [{\citenamefont {Fu}\ \emph {et~al.}(2017)\citenamefont {Fu},
  \citenamefont {Abid},\ and\ \citenamefont {Liu}}]{Fu_2017}%
  \BibitemOpen
  \bibfield  {author} {\bibinfo {author} {\bibfnamefont {B.}~\bibnamefont
  {Fu}}, \bibinfo {author} {\bibfnamefont {M.}~\bibnamefont {Abid}},\ and\
  \bibinfo {author} {\bibfnamefont {C.-C.}\ \bibnamefont {Liu}},\ }\bibfield
  {title} {\bibinfo {title} {Systematic study on stanene bulk states and the
  edge states of its zigzag nanoribbon},\ }\href
  {https://doi.org/10.1088/1367-2630/aa8c46} {\bibfield  {journal} {\bibinfo
  {journal} {New Journal of Physics}\ }\textbf {\bibinfo {volume} {19}},\
  \bibinfo {pages} {103040} (\bibinfo {year} {2017})}\BibitemShut {NoStop}%
\bibitem [{\citenamefont {Correa}\ \emph {et~al.}(2018)\citenamefont {Correa},
  \citenamefont {Pezo},\ and\ \citenamefont {Figueira}}]{Jorge18}%
  \BibitemOpen
  \bibfield  {author} {\bibinfo {author} {\bibfnamefont {J.~H.}\ \bibnamefont
  {Correa}}, \bibinfo {author} {\bibfnamefont {A.}~\bibnamefont {Pezo}},\ and\
  \bibinfo {author} {\bibfnamefont {M.~S.}\ \bibnamefont {Figueira}},\
  }\bibfield  {title} {\bibinfo {title} {Braiding of edge states in narrow
  zigzag graphene nanoribbons: Effects of third-neighbor hopping on transport
  and magnetic properties},\ }\href
  {https://doi.org/10.1103/PhysRevB.98.045419} {\bibfield  {journal} {\bibinfo
  {journal} {Phys. Rev. B}\ }\textbf {\bibinfo {volume} {98}},\ \bibinfo
  {pages} {045419} (\bibinfo {year} {2018})}\BibitemShut {NoStop}%
\bibitem [{\citenamefont {Zhu}\ \emph {et~al.}(2019)\citenamefont {Zhu},
  \citenamefont {Guo},\ and\ \citenamefont {Feng}}]{Zhu_2019}%
  \BibitemOpen
  \bibfield  {author} {\bibinfo {author} {\bibfnamefont {X.}~\bibnamefont
  {Zhu}}, \bibinfo {author} {\bibfnamefont {H.}~\bibnamefont {Guo}},\ and\
  \bibinfo {author} {\bibfnamefont {S.}~\bibnamefont {Feng}},\ }\bibfield
  {title} {\bibinfo {title} {Quantum magnetism of topologically-designed
  graphene nanoribbons},\ }\href {https://doi.org/10.1088/1361-648x/ab3f81}
  {\bibfield  {journal} {\bibinfo  {journal} {Journal of Physics: Condensed
  Matter}\ }\textbf {\bibinfo {volume} {31}},\ \bibinfo {pages} {505601}
  (\bibinfo {year} {2019})}\BibitemShut {NoStop}%
\bibitem [{\citenamefont {Xu}\ \emph {et~al.}(2012)\citenamefont {Xu},
  \citenamefont {Luo}, \citenamefont {Liu}, \citenamefont {Zheng},
  \citenamefont {Zhang}, \citenamefont {Nagase}, \citenamefont {Gao},\ and\
  \citenamefont {Lu}}]{Chengyong2012}%
  \BibitemOpen
  \bibfield  {author} {\bibinfo {author} {\bibfnamefont {C.}~\bibnamefont
  {Xu}}, \bibinfo {author} {\bibfnamefont {G.}~\bibnamefont {Luo}}, \bibinfo
  {author} {\bibfnamefont {Q.}~\bibnamefont {Liu}}, \bibinfo {author}
  {\bibfnamefont {J.}~\bibnamefont {Zheng}}, \bibinfo {author} {\bibfnamefont
  {Z.}~\bibnamefont {Zhang}}, \bibinfo {author} {\bibfnamefont
  {S.}~\bibnamefont {Nagase}}, \bibinfo {author} {\bibfnamefont
  {Z.}~\bibnamefont {Gao}},\ and\ \bibinfo {author} {\bibfnamefont
  {J.}~\bibnamefont {Lu}},\ }\bibfield  {title} {\bibinfo {title} {Giant
  magnetoresistance in silicene nanoribbons},\ }\href
  {https://doi.org/10.1039/C2NR00037G} {\bibfield  {journal} {\bibinfo
  {journal} {Nanoscale}\ }\textbf {\bibinfo {volume} {4}},\ \bibinfo {pages}
  {3111} (\bibinfo {year} {2012})}\BibitemShut {NoStop}%
\bibitem [{\citenamefont {Lü}\ \emph {et~al.}(2018)\citenamefont {Lü},
  \citenamefont {Xie},\ and\ \citenamefont {Xie}}]{Xiao_2018}%
  \BibitemOpen
  \bibfield  {author} {\bibinfo {author} {\bibfnamefont {X.~L.}\ \bibnamefont
  {Lü}}, \bibinfo {author} {\bibfnamefont {Y.}~\bibnamefont {Xie}},\ and\
  \bibinfo {author} {\bibfnamefont {H.}~\bibnamefont {Xie}},\ }\bibfield
  {title} {\bibinfo {title} {Topological and magnetic phase transition in
  silicene-like zigzag nanoribbons},\ }\href
  {https://doi.org/10.1088/1367-2630/aabc6e} {\bibfield  {journal} {\bibinfo
  {journal} {New Journal of Physics}\ }\textbf {\bibinfo {volume} {20}},\
  \bibinfo {pages} {043054} (\bibinfo {year} {2018})}\BibitemShut {NoStop}%
\bibitem [{\citenamefont {Karzig}\ \emph {et~al.}(2017)\citenamefont {Karzig},
  \citenamefont {Knapp}, \citenamefont {Lutchyn}, \citenamefont {Bonderson},
  \citenamefont {Hastings}, \citenamefont {Nayak}, \citenamefont {Alicea},
  \citenamefont {Flensberg}, \citenamefont {Plugge}, \citenamefont {Oreg},
  \citenamefont {Marcus},\ and\ \citenamefont {Freedman}}]{KarzigPRB2017}%
  \BibitemOpen
  \bibfield  {author} {\bibinfo {author} {\bibfnamefont {T.}~\bibnamefont
  {Karzig}}, \bibinfo {author} {\bibfnamefont {C.}~\bibnamefont {Knapp}},
  \bibinfo {author} {\bibfnamefont {R.~M.}\ \bibnamefont {Lutchyn}}, \bibinfo
  {author} {\bibfnamefont {P.}~\bibnamefont {Bonderson}}, \bibinfo {author}
  {\bibfnamefont {M.~B.}\ \bibnamefont {Hastings}}, \bibinfo {author}
  {\bibfnamefont {C.}~\bibnamefont {Nayak}}, \bibinfo {author} {\bibfnamefont
  {J.}~\bibnamefont {Alicea}}, \bibinfo {author} {\bibfnamefont
  {K.}~\bibnamefont {Flensberg}}, \bibinfo {author} {\bibfnamefont
  {S.}~\bibnamefont {Plugge}}, \bibinfo {author} {\bibfnamefont
  {Y.}~\bibnamefont {Oreg}}, \bibinfo {author} {\bibfnamefont {C.~M.}\
  \bibnamefont {Marcus}},\ and\ \bibinfo {author} {\bibfnamefont {M.~H.}\
  \bibnamefont {Freedman}},\ }\bibfield  {title} {\bibinfo {title} {Scalable
  designs for quasiparticle-poisoning-protected topological quantum computation
  with majorana zero modes},\ }\href
  {https://doi.org/10.1103/PhysRevB.95.235305} {\bibfield  {journal} {\bibinfo
  {journal} {Phys. Rev. B}\ }\textbf {\bibinfo {volume} {95}},\ \bibinfo
  {pages} {235305} (\bibinfo {year} {2017})}\BibitemShut {NoStop}%
\bibitem [{\citenamefont {Nayak}\ \emph {et~al.}(2008)\citenamefont {Nayak},
  \citenamefont {Simon}, \citenamefont {Stern}, \citenamefont {Freedman},\ and\
  \citenamefont {Das~Sarma}}]{NayakRMP2008}%
  \BibitemOpen
  \bibfield  {author} {\bibinfo {author} {\bibfnamefont {C.}~\bibnamefont
  {Nayak}}, \bibinfo {author} {\bibfnamefont {S.~H.}\ \bibnamefont {Simon}},
  \bibinfo {author} {\bibfnamefont {A.}~\bibnamefont {Stern}}, \bibinfo
  {author} {\bibfnamefont {M.}~\bibnamefont {Freedman}},\ and\ \bibinfo
  {author} {\bibfnamefont {S.}~\bibnamefont {Das~Sarma}},\ }\bibfield  {title}
  {\bibinfo {title} {Non-abelian anyons and topological quantum computation},\
  }\href {https://doi.org/10.1103/RevModPhys.80.1083} {\bibfield  {journal}
  {\bibinfo  {journal} {Rev. Mod. Phys.}\ }\textbf {\bibinfo {volume} {80}},\
  \bibinfo {pages} {1083} (\bibinfo {year} {2008})}\BibitemShut {NoStop}%
\bibitem [{\citenamefont {Steiner}\ and\ \citenamefont {von
  Oppen}(2020)}]{Steiner2020}%
  \BibitemOpen
  \bibfield  {author} {\bibinfo {author} {\bibfnamefont {J.~F.}\ \bibnamefont
  {Steiner}}\ and\ \bibinfo {author} {\bibfnamefont {F.}~\bibnamefont {von
  Oppen}},\ }\bibfield  {title} {\bibinfo {title} {Readout of majorana
  qubits},\ }\href {https://doi.org/10.1103/PhysRevResearch.2.033255}
  {\bibfield  {journal} {\bibinfo  {journal} {Phys. Rev. Research}\ }\textbf
  {\bibinfo {volume} {2}},\ \bibinfo {pages} {033255} (\bibinfo {year}
  {2020})}\BibitemShut {NoStop}%
\bibitem [{\citenamefont {Aasen}\ \emph {et~al.}(2016)\citenamefont {Aasen},
  \citenamefont {Hell}, \citenamefont {Mishmash}, \citenamefont {Higginbotham},
  \citenamefont {Danon}, \citenamefont {Leijnse}, \citenamefont {Jespersen},
  \citenamefont {Folk}, \citenamefont {Marcus}, \citenamefont {Flensberg},\
  and\ \citenamefont {Alicea}}]{AasenPRX2016}%
  \BibitemOpen
  \bibfield  {author} {\bibinfo {author} {\bibfnamefont {D.}~\bibnamefont
  {Aasen}}, \bibinfo {author} {\bibfnamefont {M.}~\bibnamefont {Hell}},
  \bibinfo {author} {\bibfnamefont {R.~V.}\ \bibnamefont {Mishmash}}, \bibinfo
  {author} {\bibfnamefont {A.}~\bibnamefont {Higginbotham}}, \bibinfo {author}
  {\bibfnamefont {J.}~\bibnamefont {Danon}}, \bibinfo {author} {\bibfnamefont
  {M.}~\bibnamefont {Leijnse}}, \bibinfo {author} {\bibfnamefont {T.~S.}\
  \bibnamefont {Jespersen}}, \bibinfo {author} {\bibfnamefont {J.~A.}\
  \bibnamefont {Folk}}, \bibinfo {author} {\bibfnamefont {C.~M.}\ \bibnamefont
  {Marcus}}, \bibinfo {author} {\bibfnamefont {K.}~\bibnamefont {Flensberg}},\
  and\ \bibinfo {author} {\bibfnamefont {J.}~\bibnamefont {Alicea}},\
  }\bibfield  {title} {\bibinfo {title} {Milestones toward majorana-based
  quantum computing},\ }\href {https://doi.org/10.1103/PhysRevX.6.031016}
  {\bibfield  {journal} {\bibinfo  {journal} {Phys. Rev. X}\ }\textbf {\bibinfo
  {volume} {6}},\ \bibinfo {pages} {031016} (\bibinfo {year}
  {2016})}\BibitemShut {NoStop}%
\end{thebibliography}%


\begin{thebibliography}{16}%
\makeatletter
\providecommand \@ifxundefined [1]{%
 \@ifx{#1\undefined}
}%
\providecommand \@ifnum [1]{%
 \ifnum #1\expandafter \@firstoftwo
 \else \expandafter \@secondoftwo
 \fi
}%
\providecommand \@ifx [1]{%
 \ifx #1\expandafter \@firstoftwo
 \else \expandafter \@secondoftwo
 \fi
}%
\providecommand \natexlab [1]{#1}%
\providecommand \enquote  [1]{``#1''}%
\providecommand \bibnamefont  [1]{#1}%
\providecommand \bibfnamefont [1]{#1}%
\providecommand \citenamefont [1]{#1}%
\providecommand \href@noop [0]{\@secondoftwo}%
\providecommand \href [0]{\begingroup \@sanitize@url \@href}%
\providecommand \@href[1]{\@@startlink{#1}\@@href}%
\providecommand \@@href[1]{\endgroup#1\@@endlink}%
\providecommand \@sanitize@url [0]{\catcode `\\12\catcode `\$12\catcode
  `\&12\catcode `\#12\catcode `\^12\catcode `\_12\catcode `\%12\relax}%
\providecommand \@@startlink[1]{}%
\providecommand \@@endlink[0]{}%
\providecommand \url  [0]{\begingroup\@sanitize@url \@url }%
\providecommand \@url [1]{\endgroup\@href {#1}{\urlprefix }}%
\providecommand \urlprefix  [0]{URL }%
\providecommand \Eprint [0]{\href }%
\providecommand \doibase [0]{https://doi.org/}%
\providecommand \selectlanguage [0]{\@gobble}%
\providecommand \bibinfo  [0]{\@secondoftwo}%
\providecommand \bibfield  [0]{\@secondoftwo}%
\providecommand \translation [1]{[#1]}%
\providecommand \BibitemOpen [0]{}%
\providecommand \bibitemStop [0]{}%
\providecommand \bibitemNoStop [0]{.\EOS\space}%
\providecommand \EOS [0]{\spacefactor3000\relax}%
\providecommand \BibitemShut  [1]{\csname bibitem#1\endcsname}%
\let\auto@bib@innerbib\@empty
\bibitem [{\citenamefont {Chiu}\ \emph {et~al.}(2016)\citenamefont {Chiu},
  \citenamefont {Teo}, \citenamefont {Schnyder},\ and\ \citenamefont
  {Ryu}}]{Chiu2016}%
  \BibitemOpen
  \bibfield  {author} {\bibinfo {author} {\bibfnamefont {C.-K.}\ \bibnamefont
  {Chiu}}, \bibinfo {author} {\bibfnamefont {J.~C.~Y.}\ \bibnamefont {Teo}},
  \bibinfo {author} {\bibfnamefont {A.~P.}\ \bibnamefont {Schnyder}},\ and\
  \bibinfo {author} {\bibfnamefont {S.}~\bibnamefont {Ryu}},\ }\bibfield
  {title} {\bibinfo {title} {Classification of topological quantum matter with
  symmetries},\ }\href {https://doi.org/10.1103/RevModPhys.88.035005}
  {\bibfield  {journal} {\bibinfo  {journal} {Rev. Mod. Phys.}\ }\textbf
  {\bibinfo {volume} {88}},\ \bibinfo {pages} {035005} (\bibinfo {year}
  {2016})}\BibitemShut {NoStop}%
\bibitem [{\citenamefont {Wakatsuki}\ \emph {et~al.}(2014)\citenamefont
  {Wakatsuki}, \citenamefont {Ezawa},\ and\ \citenamefont
  {Nagaosa}}]{wakatsuki2014majorana}%
  \BibitemOpen
  \bibfield  {author} {\bibinfo {author} {\bibfnamefont {R.}~\bibnamefont
  {Wakatsuki}}, \bibinfo {author} {\bibfnamefont {M.}~\bibnamefont {Ezawa}},\
  and\ \bibinfo {author} {\bibfnamefont {N.}~\bibnamefont {Nagaosa}},\
  }\bibfield  {title} {\bibinfo {title} {Majorana fermions and multiple
  topological phase transition in kitaev ladder topological superconductors},\
  }\href@noop {} {\bibfield  {journal} {\bibinfo  {journal} {Physical Review
  B}\ }\textbf {\bibinfo {volume} {89}},\ \bibinfo {pages} {174514} (\bibinfo
  {year} {2014})}\BibitemShut {NoStop}%
\bibitem [{\citenamefont {Bernevig}\ and\ \citenamefont
  {Hughes}(2013)}]{BernevigBook}%
  \BibitemOpen
  \bibfield  {author} {\bibinfo {author} {\bibfnamefont {B.~A.}\ \bibnamefont
  {Bernevig}}\ and\ \bibinfo {author} {\bibfnamefont {T.~L.}\ \bibnamefont
  {Hughes}},\ }\href {http://www.jstor.org/stable/j.ctt19cc2gc} {\emph
  {\bibinfo {title} {Topological Insulators and Topological
  Superconductors}}},\ \bibinfo {edition} {stu - student edition}\ ed.\
  (\bibinfo  {publisher} {Princeton University Press},\ \bibinfo {year}
  {2013})\BibitemShut {NoStop}%
\bibitem [{\citenamefont {Zhou}\ \emph {et~al.}(2017)\citenamefont {Zhou},
  \citenamefont {Xu},\ and\ \citenamefont {Zhou}}]{Zhou2017}%
  \BibitemOpen
  \bibfield  {author} {\bibinfo {author} {\bibfnamefont {B.-Z.}\ \bibnamefont
  {Zhou}}, \bibinfo {author} {\bibfnamefont {D.-H.}\ \bibnamefont {Xu}},\ and\
  \bibinfo {author} {\bibfnamefont {B.}~\bibnamefont {Zhou}},\ }\bibfield
  {title} {\bibinfo {title} {Majorana zero modes in a ladder of
  density-modulated kitaev superconductor chains},\ }\href
  {https://doi.org/https://doi.org/10.1016/j.physleta.2017.05.035} {\bibfield
  {journal} {\bibinfo  {journal} {Physics Letters A}\ }\textbf {\bibinfo
  {volume} {381}},\ \bibinfo {pages} {2426 } (\bibinfo {year}
  {2017})}\BibitemShut {NoStop}%
\bibitem [{\citenamefont {Maiellaro}\ \emph {et~al.}(2018)\citenamefont
  {Maiellaro}, \citenamefont {Romeo},\ and\ \citenamefont
  {Citro}}]{Maiellaro2018}%
  \BibitemOpen
  \bibfield  {author} {\bibinfo {author} {\bibfnamefont {A.}~\bibnamefont
  {Maiellaro}}, \bibinfo {author} {\bibfnamefont {F.}~\bibnamefont {Romeo}},\
  and\ \bibinfo {author} {\bibfnamefont {R.}~\bibnamefont {Citro}},\ }\bibfield
   {title} {\bibinfo {title} {Topological phase diagram of a kitaev ladder},\
  }\href {https://doi.org/10.1140/epjst/e2018-800090-y} {\bibfield  {journal}
  {\bibinfo  {journal} {The European Physical Journal Special Topics}\ }\textbf
  {\bibinfo {volume} {227}},\ \bibinfo {pages} {1397} (\bibinfo {year}
  {2018})}\BibitemShut {NoStop}%
\bibitem [{\citenamefont {Khoeini}\ \emph {et~al.}(2016)\citenamefont
  {Khoeini}, \citenamefont {Shakouri},\ and\ \citenamefont
  {Peeters}}]{Khoeini16}%
  \BibitemOpen
  \bibfield  {author} {\bibinfo {author} {\bibfnamefont {F.}~\bibnamefont
  {Khoeini}}, \bibinfo {author} {\bibfnamefont {K.}~\bibnamefont {Shakouri}},\
  and\ \bibinfo {author} {\bibfnamefont {F.~M.}\ \bibnamefont {Peeters}},\
  }\bibfield  {title} {\bibinfo {title} {Peculiar half-metallic state in zigzag
  nanoribbons of ${\mathrm{mos}}_{2}$: Spin filtering},\ }\href
  {https://doi.org/10.1103/PhysRevB.94.125412} {\bibfield  {journal} {\bibinfo
  {journal} {Phys. Rev. B}\ }\textbf {\bibinfo {volume} {94}},\ \bibinfo
  {pages} {125412} (\bibinfo {year} {2016})}\BibitemShut {NoStop}%
\bibitem [{\citenamefont {Xu}\ \emph {et~al.}(2017)\citenamefont {Xu},
  \citenamefont {Liu}, \citenamefont {Zou},\ and\ \citenamefont
  {Cheng}}]{Xu2017}%
  \BibitemOpen
  \bibfield  {author} {\bibinfo {author} {\bibfnamefont {R.}~\bibnamefont
  {Xu}}, \bibinfo {author} {\bibfnamefont {B.}~\bibnamefont {Liu}}, \bibinfo
  {author} {\bibfnamefont {X.}~\bibnamefont {Zou}},\ and\ \bibinfo {author}
  {\bibfnamefont {H.-M.}\ \bibnamefont {Cheng}},\ }\bibfield  {title} {\bibinfo
  {title} {Half-metallicity in co-doped wse2 nanoribbons},\ }\href
  {https://doi.org/10.1021/acsami.7b12196} {\bibfield  {journal} {\bibinfo
  {journal} {ACS Applied Materials \& Interfaces}\ }\textbf {\bibinfo {volume}
  {9}},\ \bibinfo {pages} {38796} (\bibinfo {year} {2017})}\BibitemShut
  {NoStop}%
\bibitem [{\citenamefont {Zhong}\ \emph {et~al.}(2015)\citenamefont {Zhong},
  \citenamefont {Quhe}, \citenamefont {Wang}, \citenamefont {Shi},\ and\
  \citenamefont {Lu}}]{Zhong_2015}%
  \BibitemOpen
  \bibfield  {author} {\bibinfo {author} {\bibfnamefont {H.-X.}\ \bibnamefont
  {Zhong}}, \bibinfo {author} {\bibfnamefont {R.-G.}\ \bibnamefont {Quhe}},
  \bibinfo {author} {\bibfnamefont {Y.-Y.}\ \bibnamefont {Wang}}, \bibinfo
  {author} {\bibfnamefont {J.-J.}\ \bibnamefont {Shi}},\ and\ \bibinfo {author}
  {\bibfnamefont {J.}~\bibnamefont {Lu}},\ }\bibfield  {title} {\bibinfo
  {title} {Silicene on substrates: A theoretical perspective},\ }\href@noop {}
  {\bibfield  {journal} {\bibinfo  {journal} {Chinese Physics B}\ }\textbf
  {\bibinfo {volume} {24}},\ \bibinfo {pages} {087308} (\bibinfo {year}
  {2015})}\BibitemShut {NoStop}%
\bibitem [{\citenamefont {Podsiad\l{}y-Paszkowska}\ and\ \citenamefont
  {Krawiec}(2015)}]{Krawiec2015}%
  \BibitemOpen
  \bibfield  {author} {\bibinfo {author} {\bibfnamefont {A.}~\bibnamefont
  {Podsiad\l{}y-Paszkowska}}\ and\ \bibinfo {author} {\bibfnamefont
  {M.}~\bibnamefont {Krawiec}},\ }\bibfield  {title} {\bibinfo {title}
  {Silicene on metallic quantum wells: An efficient way of tuning
  silicene-substrate interaction},\ }\href
  {https://doi.org/10.1103/PhysRevB.92.165411} {\bibfield  {journal} {\bibinfo
  {journal} {Phys. Rev. B}\ }\textbf {\bibinfo {volume} {92}},\ \bibinfo
  {pages} {165411} (\bibinfo {year} {2015})}\BibitemShut {NoStop}%
\bibitem [{\citenamefont {Liu}\ \emph {et~al.}(2011)\citenamefont {Liu},
  \citenamefont {Jiang},\ and\ \citenamefont {Yao}}]{Liu2011}%
  \BibitemOpen
  \bibfield  {author} {\bibinfo {author} {\bibfnamefont {C.-C.}\ \bibnamefont
  {Liu}}, \bibinfo {author} {\bibfnamefont {H.}~\bibnamefont {Jiang}},\ and\
  \bibinfo {author} {\bibfnamefont {Y.}~\bibnamefont {Yao}},\ }\bibfield
  {title} {\bibinfo {title} {Low-energy effective hamiltonian involving
  spin-orbit coupling in silicene and two-dimensional germanium and tin},\
  }\href {https://doi.org/10.1103/PhysRevB.84.195430} {\bibfield  {journal}
  {\bibinfo  {journal} {Phys. Rev. B}\ }\textbf {\bibinfo {volume} {84}},\
  \bibinfo {pages} {195430} (\bibinfo {year} {2011})}\BibitemShut {NoStop}%
\bibitem [{\citenamefont {Stepniak-Dybala}\ and\ \citenamefont
  {Krawiec}(2019)}]{Krawiec2019}%
  \BibitemOpen
  \bibfield  {author} {\bibinfo {author} {\bibfnamefont {A.}~\bibnamefont
  {Stepniak-Dybala}}\ and\ \bibinfo {author} {\bibfnamefont {M.}~\bibnamefont
  {Krawiec}},\ }\bibfield  {title} {\bibinfo {title} {Formation of silicene on
  ultrathin pb(111) films},\ }\href {https://doi.org/10.1021/acs.jpcc.9b04343}
  {\bibfield  {journal} {\bibinfo  {journal} {The Journal of Physical Chemistry
  C}\ }\textbf {\bibinfo {volume} {123}},\ \bibinfo {pages} {17019} (\bibinfo
  {year} {2019})}\BibitemShut {NoStop}%
\bibitem [{\citenamefont {Owczarek}\ and\ \citenamefont
  {Markowski}(2020)}]{Owczarek2020}%
  \BibitemOpen
  \bibfield  {author} {\bibinfo {author} {\bibfnamefont {S.}~\bibnamefont
  {Owczarek}}\ and\ \bibinfo {author} {\bibfnamefont {L.}~\bibnamefont
  {Markowski}},\ }\bibfield  {title} {\bibinfo {title} {The role of surfactant
  in two-components structures formation on si(111) surface},\ }\href
  {https://doi.org/https://doi.org/10.1016/j.susc.2019.121552} {\bibfield
  {journal} {\bibinfo  {journal} {Surface Science}\ }\textbf {\bibinfo {volume}
  {693}},\ \bibinfo {pages} {121552} (\bibinfo {year} {2020})}\BibitemShut
  {NoStop}%
\bibitem [{\citenamefont {Stepniak-Dybala}\ and\ \citenamefont
  {Krawiec}(2016)}]{Krawiec2016}%
  \BibitemOpen
  \bibfield  {author} {\bibinfo {author} {\bibfnamefont {M.}~\bibnamefont
  {Stepniak-Dybala}, \bibfnamefont {Jalochowski}}\ and\ \bibinfo {author}
  {\bibfnamefont {M.}~\bibnamefont {Krawiec}},\ }\bibfield  {title} {\bibinfo
  {title} {Silicene nanoribbons on pb-reconstructed si(111) surface},\
  }\bibfield  {journal} {\bibinfo  {journal} {Condens. Matter}\ }\textbf
  {\bibinfo {volume} {1}},\ \href {https://doi.org/10.3390/condmat1010008}
  {10.3390/condmat1010008} (\bibinfo {year} {2016})\BibitemShut {NoStop}%
\bibitem [{\citenamefont {Flensberg}\ \emph {et~al.}(2021)\citenamefont
  {Flensberg}, \citenamefont {von Oppen},\ and\ \citenamefont
  {Stern}}]{Flensberg2021}%
  \BibitemOpen
  \bibfield  {author} {\bibinfo {author} {\bibfnamefont {K.}~\bibnamefont
  {Flensberg}}, \bibinfo {author} {\bibfnamefont {F.}~\bibnamefont {von
  Oppen}},\ and\ \bibinfo {author} {\bibfnamefont {A.}~\bibnamefont {Stern}},\
  }\bibfield  {title} {\bibinfo {title} {Engineered platforms for topological
  superconductivity and majorana zero modes},\ }\bibfield  {journal} {\bibinfo
  {journal} {Nature Reviews Materials}\ }\href
  {https://doi.org/10.1038/s41578-021-00336-6} {10.1038/s41578-021-00336-6}
  (\bibinfo {year} {2021})\BibitemShut {NoStop}%
\bibitem [{\citenamefont {Ruby}\ \emph {et~al.}(2015)\citenamefont {Ruby},
  \citenamefont {Heinrich}, \citenamefont {Pascual},\ and\ \citenamefont
  {Franke}}]{Ruby_2015}%
  \BibitemOpen
  \bibfield  {author} {\bibinfo {author} {\bibfnamefont {M.}~\bibnamefont
  {Ruby}}, \bibinfo {author} {\bibfnamefont {B.~W.}\ \bibnamefont {Heinrich}},
  \bibinfo {author} {\bibfnamefont {J.~I.}\ \bibnamefont {Pascual}},\ and\
  \bibinfo {author} {\bibfnamefont {K.~J.}\ \bibnamefont {Franke}},\ }\bibfield
   {title} {\bibinfo {title} {Experimental demonstration of a two-band
  superconducting state for lead using scanning tunneling spectroscopy},\
  }\href {https://doi.org/10.1103/PhysRevLett.114.157001} {\bibfield  {journal}
  {\bibinfo  {journal} {Phys. Rev. Lett.}\ }\textbf {\bibinfo {volume} {114}},\
  \bibinfo {pages} {157001} (\bibinfo {year} {2015})}\BibitemShut {NoStop}%
\bibitem [{\citenamefont {Persson}(2018)}]{Persson_2018}%
  \BibitemOpen
  \bibfield  {author} {\bibinfo {author} {\bibfnamefont {J.~R.}\ \bibnamefont
  {Persson}},\ }\bibfield  {title} {\bibinfo {title} {Hyperfine structure and
  hyperfine anomaly in pb},\ }\href {https://doi.org/10.1088/2399-6528/aac52b}
  {\bibfield  {journal} {\bibinfo  {journal} {Journal of Physics
  Communications}\ }\textbf {\bibinfo {volume} {2}},\ \bibinfo {pages} {055028}
  (\bibinfo {year} {2018})}\BibitemShut {NoStop}%
\end{thebibliography}%

\end{document}



\title{Supplemental Material: Spin-polarized Majorana zero-modes in double zigzag honeycomb nanoribbons}

\author{R. C. Bento Ribeiro}
\affiliation{Instituto de F\'{i}sica, Universidade Federal Fluminense, Av. Litor\^anea s/N, CEP: 24210-340, Niter\'oi, RJ, Brazil}
\affiliation{Centro Brasileiro de Pesquisas Físicas, Rua Dr. Xavier Sigaud, 150, Urca 22290-180, Rio de Janeiro, RJ, Brazil}
\author{J. H. Correa}
\affiliation{Universidad Tecnol\'ogica del Per\'{u},  Nathalio S\'anchez, 125, Lima, Per\'{u}}
\affiliation{Instituto de Física, Universidade de Brasília, Brasília-DF 70919-970, Brazil} 

\author{L. S. Ricco}
\affiliation{Science Institute, University of Iceland, Dunhagi-3, IS-107,Reykjavik, Iceland}


\author{A. C. Seridonio}
\affiliation{S\~ao Paulo State University (Unesp), School of Engineering, Department of Physics and Chemistry, 15385-000, Ilha Solteira-SP, Brazil}
\affiliation{S\~ao Paulo State University (Unesp), IGCE, Department of Physics, 13506-970, Rio Claro-SP, Brazil}

\author{M. S. Figueira}
\email[corresponding author:]{figueira7255@gmail.com}
\affiliation{Instituto de F\'{i}sica, Universidade Federal Fluminense, Av. Litor\^anea s/N, CEP: 24210-340, Niter\'oi, RJ, Brazil}

\date{\today}
\maketitle

\section{Note $1$ - Infinite double KzHNR}

We first consider the spinless case. This section calculates the winding numbers and the band structure by considering the Fourier transform of the Hamiltonian $H$ [Eq.~(2-3) in the main text]. The total Hamiltonian is given by
\begin{equation}
H=H_{t}+H_{\Delta} ,
\label{Htot}
\end{equation}
with the first nearest neighbor hopping and the superconductor pairing term. In the momentum representation, the Hamiltonian can be written as
\begin{equation}
\begin{split} 
H_{t}=- \sum_{k,n}[ \mu (a_{k,n}^{\dagger}a_{k,n} + b_{k,n}^{\dagger}b_{k,n})+
t(a_{k,n}^{\dagger}b_{k,n-1} - 
2a_{k,n}^{\dagger}b_{k,n}\cos(ka/2)) +H.c.] , 
\end{split}
\label{Firstspin}
\end{equation}
\begin{equation} 
\begin{split}
H_{\Delta}=\sum_{k,n} [2i\Delta\sin(k)(a_{k,n}^{\dagger}a_{-k,n}^{\dagger}+
b_{k,n+1}^{\dagger}b_{-k,n+1}^{\dagger}) +H.c .] , 
\end{split}
\label{pairingspin}
\end{equation}
where $n = 1, 2$, corresponding to the top and bottom chain index.   In the  Bogoliubov-de Gennes (BdG) form we can express the Hamiltonian as $\frac{1}{2}\sum_{k} \Psi^{\dagger}h(k)\Psi$, with  $\Psi \equiv  \left( \begin{array}{cccccccccccccccccccccccccccccccccccccccccc}  a_{k,1 }, a_{-k,1}^{\dagger},  b_{k,1},b_{-k,1}^{\dagger}, a_{k,2 }, a_{-k,2}^{\dagger}, b_{k,2},  b_{-k,2}^{\dagger}  
     \end{array} \right)$. We obtain the total Hamiltonian in the following matrix form
\begin{equation}
     \begin{aligned}
     h(k)=  \left( \begin{array}{cccccccc}
-2 \mu &  \Delta_k  & -\epsilon  & 0& 0 &  0 &  0  &  0\\
\Delta_k *  & + 2 \mu & 0 & \epsilon & 0 &  0 &  0 &  0\\
-\epsilon& 0 &  -2 \mu &\Delta_k   & -t &  0 &  0 &  0  \\
0& \epsilon & \Delta_k *  & +2 \mu  & 0  &  t &  0 &  0  \\
0 &0 &  -t & 0 &  -2 \mu &  \Delta_k &  -\epsilon &  0 \\
0 &0 & 0 & t&\Delta_k * &  + 2 \mu &  0 &  \epsilon \\
0 &0 &0  & 0 &  -\epsilon & 0 &   -2 \mu & \Delta_k\\
 0 &  0& 0 & 0  &  0 &  \epsilon & \Delta_k * &  +2 \mu \\
    \end{array} \right) ,
    \end{aligned}
    \label{hamiltonian_matriz}
    \end{equation}
where  $\Delta_k= 2i\Delta\sin(k)$ e $\epsilon =-2t\cos(k/2)$. This Hamiltonian has the dispersion relations  given by 
\begin{equation}
\begin{split}
&E_{1,2,3,4}=\pm\sqrt{\frac{-2\Delta^{2}+2\zeta_{1}+t^{2}\pm\tilde{\epsilon}(4\mu+t)}{2},}\,\,\,\, E_{5,6}=\pm\sqrt{\frac{-2\Delta^{2}+2\zeta_{2}+t^{2}-\tilde{\epsilon}(4\mu-t)}{2}} ,\\ 
&\text{and}\,\,\,\,E_{7,8}=\pm\sqrt{-\Delta^{2}+\epsilon^{2}+2\zeta_{3}+\tilde{\epsilon}(2\mu-t/2)},\label{E_Dispersion}
\end{split}    
\end{equation}
with $\zeta_{1}=\epsilon^{2}+ 2\mu^{2} +2\mu t$, $\zeta_{2}=\epsilon^{2}+ 4\mu^{2} -2\mu t$, $\zeta_{3}=2\mu^{2} -\mu t + t^{2}$ and ${\tilde{\epsilon}=\sqrt{4\epsilon^{2}+ t^{2}}}$. It is worth noting that Eq.~(\ref{hamiltonian_matriz}) satisfies both the particle-hole and time-reversal symmetries, since
\begin{equation}
\mathcal{C}h(k)\mathcal{C}^{-1}=-h(-k) ,
\label{charge_conjugation}   
\end{equation}
and 
\begin{equation}
\mathcal{T}h(k) \mathcal{T}^{-1} = h(-k),\label{time_reversal} 
\end{equation}
where $\mathcal{C}$ and $\mathcal{T}$ are charge conjugation and time-reversal operators~\cite{Chiu2016,wakatsuki2014majorana}, respectively. Eq.~(\ref{hamiltonian_matriz}) also satisfies the chiral symmetry
\begin{equation}
\mathcal{K}h(k)\mathcal{K}^{-1} = -h(k), \label{Chiral}    
\end{equation}
in which the chiral operator is defined by the anti-commutation relation $[\mathcal{K},h(k)]_{+}=0$. Thus, one can write $h(k)$ in its corresponding chiral form by performing the following unitary transformation:
\begin{equation}
 \tilde{h}(k) = \mathcal{U}^{\dagger}h(k)\mathcal{U} = \begin{bmatrix}0 & A(k)\\
A^{*}(k) & 0
\end{bmatrix},  \label{ChiralForm}
\end{equation}
where $A(k)$ is a $4 \times 4$ chiral matrix given by
\begin{align}
    A(k)=\displaystyle \left[\begin{matrix}- 4 i \Delta \sin{\left(k \right)} + 4 \mu & - 4 t \cos{\left(\frac{k}{2} \right)} & 0 & 0\\- 4 t \cos{\left(\frac{k}{2} \right)} & - 4 i \Delta \sin{\left(k \right)} + 4 \mu & 2 t & 0\\0 & 2 t & - 4 i \Delta \sin{\left(k \right)} + 4 \mu & - 4 t \cos{\left(\frac{k}{2} \right)}\\0 & 0 & - 4 t \cos{\left(\frac{k}{2} \right)} & - 4 i \Delta \sin{\left(k \right)} + 4 \mu\end{matrix}\right].\label{A(k)}
\end{align}
Once particle-hole, time-reversal, and chiral symmetries are preserved by $h(k)$, the corresponding system belongs to the BDI symmetry group class with $\mathbb{Z}$ index~\cite{Chiu2016,wakatsuki2014majorana}, with its topology being characterized by the associated Chern number invariant~\cite{BernevigBook}, i.e., the winding number~\cite{Zhou2017,Maiellaro2018}
\begin{equation}
    \begin{split}
          W= Tr \int_{-\pi}^{\pi} \frac{dk}{2 \pi i} A_{k} ^{-1} \partial _k A_{k} = - \int_{-\pi}^{\pi} \frac{dk}{2 \pi i} \partial _k ln[Det(A_k)], 
    \end{split}
\end{equation}
which gives the number of MZMs at the edges of the spinless KzHNR, as discussed in Fig.~2 of the main text.

\begin{figure}[t]
 \centerline{\includegraphics[width=7.0in,keepaspectratio]{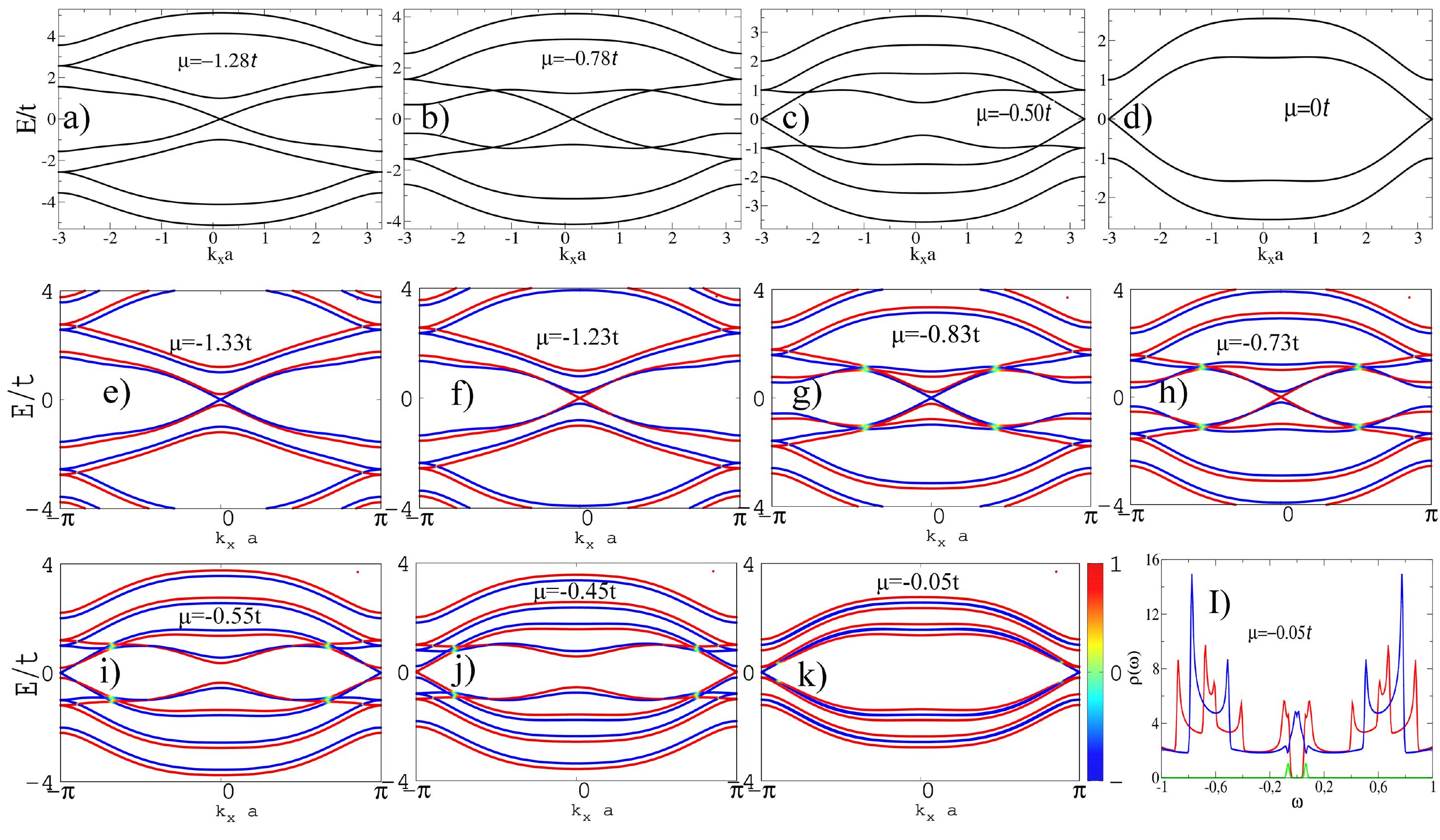}}\caption{Panels (a)-(d): band structure for a $ N=2$ infinity double-spinless KzHNR, considering $\mu$ values where the TPTs occur. Panels (e)-(k): the same as top panels, but considering both spin components (infinite double-spin KzHNR). The energies are measured in units of $t$ as described in the main text, and we fix the \textit{p}-wave superconducting pairing parameter $\Delta = 0.5t$. Additionally, the parameters of the double-spin case (middle and bottom panels) are $\lambda_R=0.05t$ and $V_z=0.1t$. In the double-spinless case we have four TPTs occurring at: a)$\mu=-1.28t$ b) $\mu=-0.78t$ c) $\mu=-0.50t$ d) $\mu=0$. For the double-spin case, the TPTs occur at e)$\mu=-1.33t$ with spin down, f)$\mu=-1.23t$ with spin up g)$\mu=-0.84t$ with spin down, h)$\mu=-0.73t$ with spin up, i)$\mu=-0.55t$ with spin down j) $\mu=-0.45t$ with spin up k)$\mu=-0.05t$ with spin down. The density of states, corresponding to the band structure depicted in (k), is plotted in (l), where we can observe the half-metallicity characteristic of those TPTs.}\label{fig:three graphs}.
\end{figure}

In Fig.\ref{fig:three graphs} (a)-(d), we plot the band relation dispersion [Eq.~(\ref{E_Dispersion})] for a $N=2$ infinity double-spinless KzHNR, considering $\mu$ values where the TPTs occur. For a) $ \mu = − 1.28t $ and b) $ \mu = − 0.78t $, the gap closes at $ k = 0 $, defining two TPTs and forming a topological phase in this interval with winding number equal  to $W=1$.  A new TPT occurs at c) $\mu=-0.50t$, with the gap closing at 
$\vec{k}= \pi$, and defining a new topological region with winding number equal to $W=2$  between $ \mu = −0.78t$  and $\mu = −0.50t$. Finally, at d) $\mu=0$ the gap closes again at $\vec{k}= \pi$, defining another TPT, and forming a topological phase with winding number equal to $W=1$ between $ \mu = −0.50t$  and $ \mu = 0$. The same transitions appears at the same values for positive chemical potentials, once the system described by Eq.~(\ref{E_Dispersion}) exhibits a particle-hole symmetry.

Considering the Fourier transform of the total Hamiltonian (Eq. 4 of the main text), we calculate the topological phase transitions of the infinite spinfull case as a function of the chemical potential. Again we consider $n = 1, 2$, corresponding to the top and bottom chain index.

\begin{equation}
\begin{split} 
H_{t}=-\sum^{N}_{k,n,\sigma}  \left[\mu (a_{k,n,\sigma}^{\dagger}a_{k,n,\sigma} + b_{k,n,\sigma}^{\dagger}b_{k,n,\sigma})+t(a_{k,n,\sigma}^{\dagger}b_{k,n-1,\sigma} - 
2a_{k,n,\sigma}^{\dagger}b_{k,n,\sigma}\cos(ka/2)) +H.c. \right] , 
\end{split}
\label{Firstspin}
\end{equation}
\begin{equation} 
\begin{split}
H_{\Delta}=\sum^{N}_{k,n,\sigma} \left[\Delta_{k}(a_{k,n,\sigma}^{\dagger}a_{-k,n,\sigma}^{\dagger}+b_{k,n+1,\sigma}^{\dagger}b_{-k,n+1,\sigma}^{\dagger}) +H.c. \right] \hspace{0.3cm}-\hspace{0.3cm} \Delta_{k}=2i\Delta\sin(k) ,
\end{split}
\label{pairingspin}
\end{equation}
\begin{eqnarray}
\begin{split}
    &H_R  = \sum^{N}_{k,n,\sigma}\Big[\Lambda_{R\sigma}\Big(-a_{k,n,\sigma}^{\dagger}b_{k,n-1, \overline{\sigma}}-
		 2\cos{\left( k/2-\frac{2\pi}{3}\right)}a_{k,n,\sigma}^{\dagger}b_{k,n, \overline{\sigma}} 
    + b_{k,n,\sigma}^{\dagger}a_{k,n+1, \overline{\sigma}}+ \\
		&2\cos{\left(k/2+\frac{2\pi}{3}\right)}b_{k,n,\sigma}^{\dagger}a_{k,n, \overline{\sigma}}\Big) +H.c. \Big]  \hspace{0.3cm}-\hspace{0.3cm} \Lambda_{R\sigma}=i\lambda_{R}sign(\sigma) ,
    \end{split}
		\label{Rashba}
\end{eqnarray}
\begin{eqnarray}
H_{z}  =  \sum^N_{k,n,\sigma} sign(\sigma)\lambda_{z}(a_{k,n,\sigma}^{\dagger}a_{k,n,\sigma} + b_{k,n,\sigma}^{\dagger}b_{k,n,\sigma}+H.c.) , 
\label{zeeman}
\end{eqnarray}
In Fig.\ref{fig:three graphs}(e)-(k), we plot the band structure [Eqs.~(\ref{Firstspin}) and (\ref{zeeman})] for a $ N=2$ infinity double-spin KzHNR, considering $\mu$ values where the TPTs occur. To discriminate the two possible spin orientations in the edges of the KzHNR, we  introduce two new physical effects: the extrinsic RSOC and the EMF given by Eqs.~(\ref{Rashba}) and Eq.~(\ref{zeeman}). When compared to the earlier case, each transition splits into two, one with spin up and the other with spin down, where the split is tuned by the $\lambda_{R}=\pm 0.05t$ parameter. For example, the TPT that occurs at Fig~\ref{fig:three graphs}(a) $\mu=-1.28t$, for the spinless case splits into $\mu \rightarrow \mu \pm \lambda_{R}=-1.33t; -1.23t$ in Figs~\ref{fig:three graphs}(e) and (f), respectively. However, in the double-spin case, we did not obtain the winding numbers due to the involved complexity of the calculations.

In Fig.~\ref{fig:three graphs}(l), we plot a typical density of states for $\mu=-0.05t$ in a point where a TPT occurs and that exhibits half-metallicity; what is another striking characteristic that occurs in all the other Majorana TPTs of the system. This effect leads the double-spin KzHNR into a half-metallic state as indicated in Fig. \ref{fig:three graphs}(l), resulting in insulating behavior for one spin component and metallic behavior for the other component \cite{Khoeini16, Xu2017}.

\section{Note $2$ - Finite double-spin KzHNR: Parameter analysis}

\begin{figure}[h]
	\centerline{\includegraphics[width=3.0in,keepaspectratio]{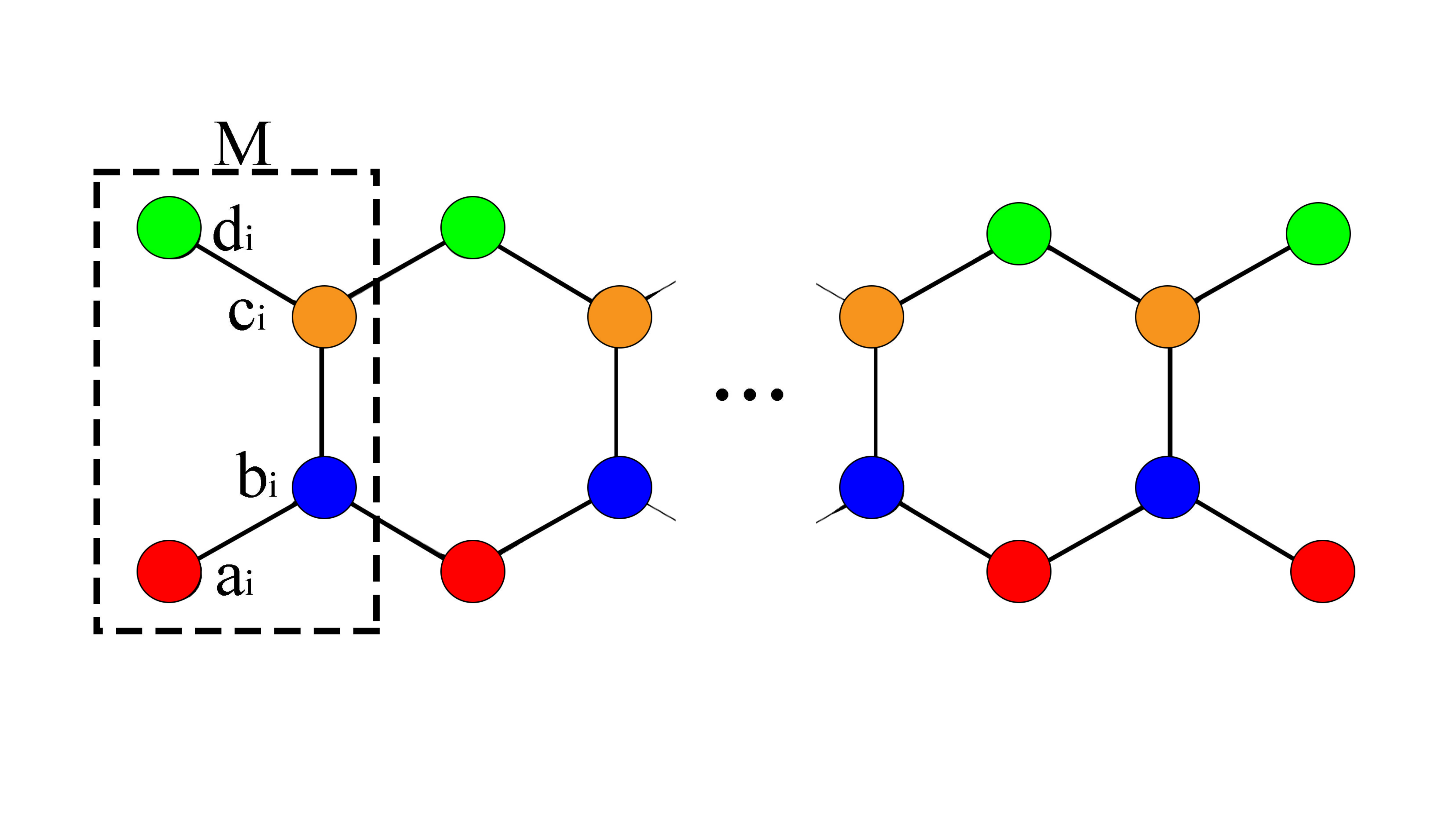}}\caption{Scheme of the basis adopted in the Hamiltonian given by the Eqs. Eqs.~(4-8) of the main text describes the finite double-spin KzHNR chain of width $N=2$. The unit cell represented by the dashed rectangular area is built with four distinct atoms $a_{i}$, $b_{i}$, $c_{i}$ and $d_{i}$.
	\label{fig:SchemeNewBasis}}
\end{figure}    
To describe the double-spin KzHNR of finite length, we employ the Hamiltonian given by Eqs.~(4-8) of the main text, which is solved numerically using the basis represented in Fig.~\ref{fig:SchemeNewBasis}. The calculation becomes more time-consuming as the Hamiltonian matrix dimension grows with the value of $M$.

\begin{figure}[h]
\centerline{\includegraphics[width=7.0in,keepaspectratio]{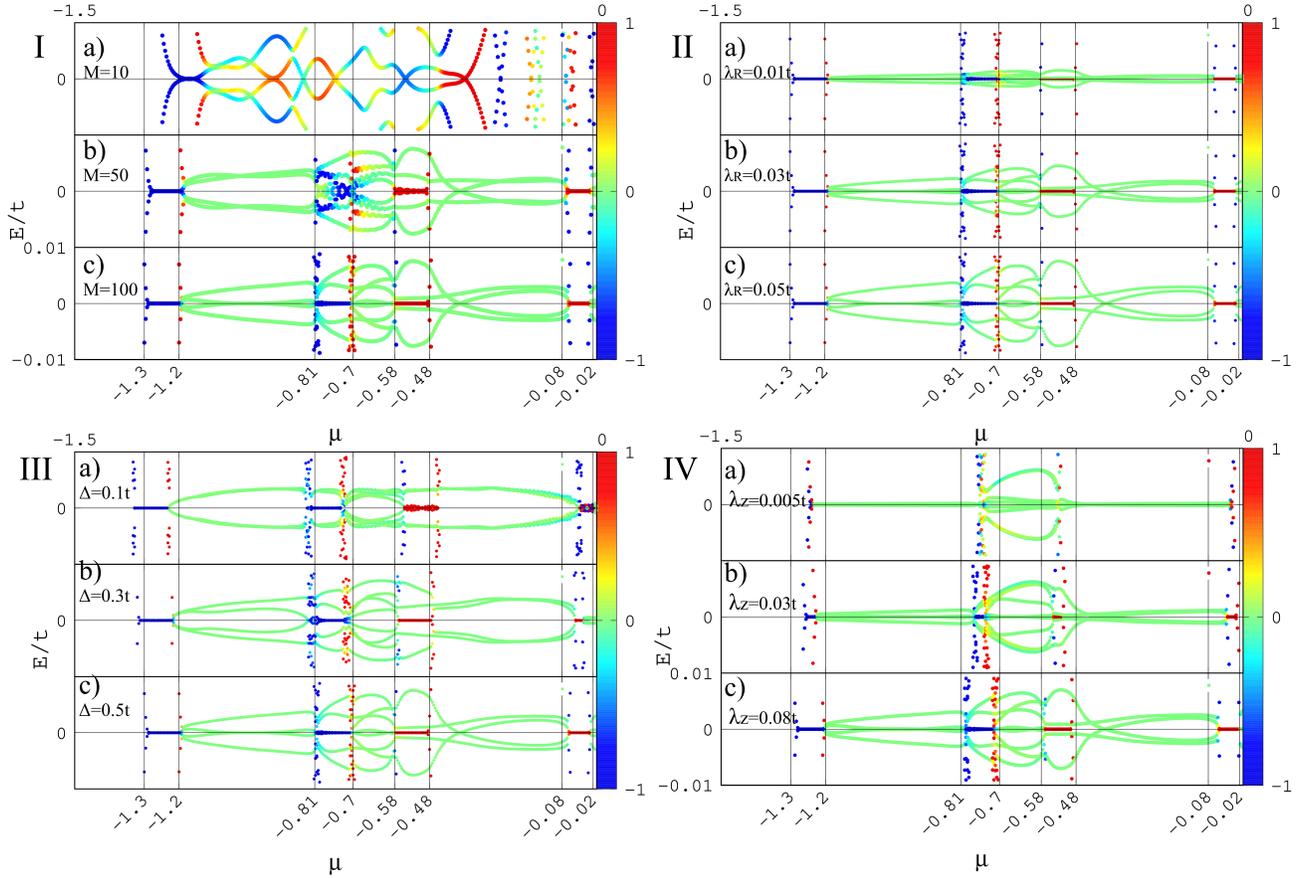}}
\caption{Formation of MZMs parameter study: Energy spectra for a $N=2$ finite double-spin KzHNR as a function of $\mu$. We employed the same parameters set used in the simulations of the main paper (cf. figure 3 of the main text): $\Delta = 0.5t$,  $\lambda_R=0.05t$ and $V_{z}=0.1t$,  but only changing the particular parameter indicated in the figure: I)Length: $M=10, 50, 100$. II) Extrinsic Rashba spin-orbit coupling $\lambda_{R}=0.01t, 0.03t, 0.05t.$ III) Superconductor pairing $\Delta=0.1t, 0.3t, 0.5t$. IV) EMF $\lambda_{Z}=0.005t, 0.03, 0.08t$. 
	\label{fig:Parameter_study}}
\end{figure}

In Figs.~\ref{fig:Parameter_study}(I)-(IV), we plot the energy spectra as a function of $\mu$ of a double-spin KzHNR for distinct parameters of the Hamiltonian. As in the main text, we keep $\Delta = 0.5t$, $\lambda_R=0.05t$, and $V_{z}=0.1t$ except the corresponding varied parameter. We explicitly indicate it in the panels.

Figure \ref{fig:Parameter_study}-I shows the dependence of the energy spectra as a function of $\mu$ for lengths $M=10$(a), $50$(b) and $100$(c) of the double-spin KzHNR, respectively. In panel (a), we can verify oscillatory patterns for the smallest double spin-KzHNR structure. The oscillating behavior is expected to appear in small Majorana nanowires due to overlapping MZMs of opposite edges. For $M=50$ (b), we observe the appearance of MZMs at the real axis around $\mu=0$ and in the inferior band region in the range, $\mu=[-1.2,-1.3]$. Panel (c) depicts the case of $M=100$, showing the MZMs on the real axis in all the available topological regions.

Figure \ref{fig:Parameter_study}-II shows the dependence of the energy spectra as a function of $\mu$ with the extrinsic RSOC parameters $\lambda_{R}=0.01t$, $0.03t$ and $0.05t$. The result shows that a low value of $\lambda_{R}$ is sufficient to generate well-defined MZMs on the real axis in all the topological regions.

In Fig.~\ref{fig:Parameter_study}-III, we observe the dependence of the energy spectra as a function of $\mu$ for $\Delta=0.1t$, $0.3t$ and $0.5t$, in panels (a), (b) and (c), respectively. These profiles indicate that the parameter p-wave paring $\Delta$ strongly affects the MZMs formation on the real axis. The MZMs are formed first, for $\Delta=0.1t$ (a) in the range $\mu=[-1.2,-1.3]$. Only when $\Delta=0.5t$ (b), the MZMs emerges at around $\mu=0$. Well-defined MZMs arises in the region of $\mu \simeq 0$ only for higher values of $\Delta$.

Fig.~\ref{fig:Parameter_study}-IV shows the dependence of the energy spectra with the EMF for
$\lambda_{Z}=0005t$(a), $003t$(b) and $008t$(c). The EMF acts uniformly over the MZMs formation for all $\mu$ values. The enhancement of $\lambda_{Z}$ also increases the number of MZMs over the real axis.


\section{Note $3$ - Experimental prospects about a double-spin KzHNR based on silicene layers deposited on top of a lead superconducting substrate}

Silicene layers were grown on Ag, Ir, Ca, and Pb, among other metallic substrates~\cite{Zhong_2015}. However, due to the considerable band hybridization between silicene and the substrate, the absence of Dirac cones is a characteristic feature, except for Ca and Pb, which preserves the Dirac cone below the Fermi surface~\cite{Krawiec2015}. Interesting enough, Pb is a conventional Bardeen-Cooper-Schrieffer (BCS) superconductor, with a relatively high critical temperature of $T_{C}=\SI{7.2}{\kelvin}$ and a strong intrinsic RSOC around $\SI{1}{\milli\electronvolt}$~\cite{Liu2011}. Using scanning tunneling microscopy and spectroscopy supported by ab initio density functional theory calculations, some successful experiments have revealed the epitaxial growth of silicene on top of the ultrathin Pb(111) films and $\sqrt{3} \times \sqrt{3}$ phase of silicene in a Pb(111) slab, additionally covered by a single Pb layer~\cite{Krawiec2019,Owczarek2020,Krawiec2016}. The stability of silicene nanoribbons was achieved by passivation of the bare Si(111) surface by Pb atoms, which lowers the surface energy.

Due to its buckled structure and large silicon ionic radius, silicene has a relatively large effective intrinsic spin-orbit (ISOC) gap of $\SI{1.55}{\milli\electronvolt}$, and an intrinsic RSOC of $\SI{0.7}{\milli\electronvolt}$~\cite{Liu2011}. However, this intrinsic RSOC is not enough to break the spin degeneracy of the energy spectra.
Based on the Ref.~\onlinecite{Flensberg2021}, we estimate the corresponding parameters for a silicene double-spin KzHNR structure on top of a Pb superconducting substrate.

To develop a topological phase in the silicene-lead superconducting hybrid structure, the Zeeman energy $\lambda_{Z}$ should satisfy the relation $\lambda_{Z}>\sqrt{\Delta_{ind}^{2} + \mu^{2}}$, where $\Delta_{ind}^{2}$ is the induced superconducting gap in the silicene layers due to proximity effect and $\lambda_{Z}=\frac{1}{2}g_{Si}\mu_{B}B$, with the silicene g-factor $g_{Si}$ and the Bohr magneton $\mu_{B}$.
On the other hand, the Pb superconductor quasiparticle excitation spectrum exhibits a gap of width $E_{g}=2\Delta_{S} \simeq \SI{2.73}{\milli\electronvolt}$ around the Fermi level~\cite{Ruby_2015}, with $\Delta_{S}$  being the binding energy of the Cooper pairs. Thus, $\lambda_{Z}$ should not assume values bigger than Pb substrate superconducting energy gap, i.e, $\frac{1}{2}g_{Pb}\mu_{B}B < E_{g}$. Considering $g_{Si} \simeq 2.0$ associated to the free-electron silicene conduction band and $g_{Pb}=1.5$, associated to the $ \ce{^{3}}P_{1}$ configuration of Pb~\cite{Persson_2018}, we estimate that $\SI{3.64}{\milli\electronvolt} > \sqrt{\Delta_{ind}^{2} + \mu^{2}}$, which turns our proposal feasible from the experimental point of view.

\bibliography{references}